\documentclass[12pt]{article}

\usepackage{amssymb,amsmath,amsfonts,array,eurosym,geometry,ulem,graphicx,color,setspace, comment,footmisc,caption,natbib,pdflscape,array, graphicx, float, booktabs, appendix, tabularx, mathtools, threeparttable, rotating, lscape, enumerate, bbm, multirow,enumitem,subcaption,arydshln,tikz}

\usetikzlibrary{positioning, decorations.markings,decorations.pathreplacing,calligraphy}
\usepackage{hyperref}
\hypersetup{
colorlinks=true,
citecolor=blue,
linkcolor=blue,
filecolor=blue,      
urlcolor=blue,
}

\newcolumntype{L}[1]{>{\raggedright\let\newline\\arraybackslash\hspace{0pt}}m{#1}}
\newcolumntype{C}[1]{>{\centering\let\newline\\arraybackslash\hspace{0pt}}m{#1}}
\newcolumntype{R}[1]{>{\raggedleft\let\newline\\arraybackslash\hspace{0pt}}m{#1}}
\def\sym#1{\ifmmode^{#1}\else\(^{#1}\)\fi}
\DeclareMathOperator{\E}{\mathbb{E}}

\begin{document}

\begin{titlepage}
\title{The heterogeneous impact of the EU-Canada\\
\vspace{-0.7em}
agreement with causal machine learning\thanks{\footnotesize{We thank participants at the European Winter Meeting of the Econometric Society 2024. We are grateful for the useful comments by Gabor Bekes, Giovanna D'Inverno, Daniela Maggioni, and John Morrow. Lionel Fontagn\'e thanks the support of the EUR grant ANR-17-EURE-0001 and the Institute for Macroeconomic and International policies, i-MIP. The views expressed in this paper are the authors'. Armando Rungi is grateful for the support of a PRIN grant CUP: D53D23006740006, funded by the European Union – Next Generation EU.}}}
\author{Lionel Fontagn\'e \thanks{\scriptsize{Mail to: \href{mailto://lionel.fontagne@univ-paris1.fr}{\color{blue}lionel.fontagne@psemail.eu}. Paris School of Economics, 48 boulevard Jourdan 75014 Paris. }}\and Francesca Micocci\thanks{\scriptsize{Mail to: \href{mailto://francesca.micocci@imtlucca.it}{\color{blue}francesca.micocci@imtlucca.it}. Laboratory for the Analysis of Complex Economic Systems, IMT School for Advanced Studies, piazza San Francesco 19 - 55100 Lucca, Italy.} } \and Armando Rungi\thanks{\scriptsize Mail to: \href{mailto://armando.rungi@imtlucca.it}{\color{blue}armando.rungi@imtlucca.it}. Laboratory for the Analysis of Complex Economic Systems, IMT School for Advanced Studies, piazza San Francesco 19 - 55100 Lucca, Italy.}}

\maketitle
\vspace{-2.5em}

\begin{abstract}
\footnotesize
\singlespacing
This paper introduces a causal machine learning approach to investigate the effects of free trade agreements and applies it to the EU-Canada Comprehensive Economic and Trade Agreement (CETA). Previous estimates of the impact of trade liberalization have been found to be unstable and contradictory, possibly due to the presence of heterogeneous treatment effects. The matrix completion estimator computes multidimensional counterfactuals in trade data at the firm, product, and destination levels. Compared with other estimators, it relies on a weaker exogeneity assumption and a more general functional form. In the case of CETA, we obtain both positive and negative idiosyncratic treatment effects at the product-destination level, although the sales-weighted average treatment effect is 6.4\% in the year after the agreement. At the same time, we can estimate idiosyncratic treatment effects for the extensive margin at the product-destination level; thus, we find product churning beyond regular entry-exit dynamics: 8.1\% that were not previously exported, and about 7.3\% that are no longer exported. Finally, we consider the case of multiproduct firms after ranking product portfolios. After CETA, we observe a reallocation of French exports toward the first and most exported products, possibly driven by increased competition in the local market by other European producers after trade liberalization.

\vspace{20pt}
\bigskip
\end{abstract}
\setcounter{page}{0}
\thispagestyle{empty}
\end{titlepage}
\pagebreak \newpage

\onehalfspacing
\section{Introduction}\label{sec: introduction} 
Ex-post estimates of the impact of Free Trade Agreements (FTAs) have been shown
to be both unstable and fragile \citep{Baier_et_al_2019}. This can primarily be attributed to the challenges of effectively addressing endogenous selection in trade agreements and the design of sensible counterfactuals. Due to the phasing-in of tariff reductions, staggered treatment adoption is often an issue raised when evaluating trade agreements \citep{yotov_2024}. Even if the design is not staggered, ``forbidden comparisons'' can be problematic if the treatment is not binary \citep{chaisemartin_2023}. These empirical challenges are further aggravated by the presence of heterogeneous firms, which can sell multiple products and operate across multiple destinations.

In this contribution, we propose a causal machine-learning approach to evaluate FTAs. In particular, we adopt a matrix completion algorithm for panel data, originally proposed by \citet{athey2021matrix}, to uncover the impact of the  CETA (EU-Canada Comprehensive Economic and Trade Agreement) on French trade, using monthly customs data on the universe of French exports. Our approach allows multidimensional counterfactuals, a weaker exogeneity assumption, and a more flexible functional form. 

Notably, machine learning methods are increasingly used in economics for prediction and classification \citep{Mullainathan_Spiess_2017, Athey_Imbens_2019, Breinlich_et_al_2022, Bartelme_et_al_2024}, but also for causal inference \citep{Athey_2015, Knaus_2020, Baiardi_Naghi_2024}. Similar to synthetic control methods, the matrix completion method relaxes the assumption of parallel trends by allowing for unobserved time-varying confounders \citet{Xu_2017,Poulos_et_al_2021}. It uses observed variation in adjacent outcomes and covariates, combined with the absence of linear functional forms, to allow units of observation to follow different paths. Unlike a difference-in-difference estimator, and similar to synthetic control, it does not compare treated units with a control group. According to the weaker exogeneity assumption, the errors in the predictive equations for each counterfactual observation are conditionally independent, conditionally mean zero, and conditionally independent of the treatment \citet{Xu_2017}.

Specifically, in our exercise, we derive a matrix of observed outcomes from the French customs data to be partitioned into: i) treated vs. untreated observations, depending on whether the units of observation (products or firms) have experienced a reduction in tariffs or a change in quotas thanks to the CETA; ii) observations before and after the signing of the CETA.

Crucially, we can follow the application of the CETA agreement with monthly trade data. As the signature occurred in September 2017, we split the timeline around that threshold. Then, we first perform our exercise at the product-destination level in the period starting from 2012M9-2013M08, considering as treated the products included in CETA; therefore, we proceed at the firm level, considering the case of multiproduct firms and their portfolio ranking based on the information in the two years before the treaty agreement. In the latter case, we consider a firm as treated when at least one of its products is listed in the treaty. To reduce matrix sparsity and an excessive number of zeros, we aggregate fifteen alternative top destinations and close the matrix with a rest-of-the-world category.  In the product-level case, the trade matrix in each period has cells for 5,118 products at the HS 6-digit level, 16 alternative destinations (including Canada), and 6 years. In the second case, the matrix identifies, for each period, 3,791 multiproduct firms, and up to 3 of their most exported products. 

Once the trade matrix is constructed, we can drop the observations for treated units, i.e., those whose trade regime changed after the agreement entered into force, and thus predict their values as if the CETA had not been signed. Predictions are obtained by reducing the original trade matrix to a low-rank structure via nuclear norm regularization.

To evaluate the predictive power of matrix completion, we follow standard machine learning approaches: we train the algorithm on five random folds on the part of the matrix containing untreated units, and then assess out-of-sample how far predictions are from the observed outcomes in the same portion of the matrix with no treated observations. If the predictive power is good enough, we assume the algorithm also performs well on the portion of the matrix with treated units, where the predictions represent counterfactuals.

For our purpose, CETA is a compelling example of an FTA whose negotiation has been intricate, lengthy, and complex. It took 10 years from the first discussions\footnote{It dates back to a Canada-EU bilateral summit in Berlin in 2007.} to the agreement's provisional entry into force in 2017. According to its provisional enforcement, most of the trade provisions in the agreement have already been applied, although it is still awaiting final ratification by all EU members.\footnote{Even if the European Commission is solely in the competence of the trade policy of the European Union, in  July 2016 it was decided that CETA qualified as a \textit{mixed agreement} because it touches upon other policy domains different from trade, and thus it needed to be ratified through national procedures.} During the negotiations, France emerged as one of the main proponents of establishing a closer trading relationship with Canada. A shared colonial past, a common language,\footnote{English and French have been established as joint official languages since 1969.}, and similar consumer preferences give Canada more than an incentive to trade with France. Ratification by the French Assembly was voted on in July 2019, and the agreement was examined and eventually rejected by the French Senate in March 2024. 

Yet, an asymmetry was evident from the beginning for all parties involved in the negotiation. The treaty would have \textit{prima facie} been more relevant for Canada than for European countries. However, the EU interest was to foster unprecedented economic cooperation with new partners after the rise of emerging markets like China \citep{hubner2017eu} and to have a testing ground for \textit{deep trade agreements} covering areas beyond tariffs. Notably, an asymmetry in the size of the parties involved in the Treaty makes local competition among European exporters potentially larger than the relatively smaller positive demand shock induced by trade liberalization.\footnote{Please note that Canada's GDP is similar in size to Italy's. France is Canada's ninth-largest trading partner and the fourth-largest among EU members. At the same time, Canada ranks only 30th among trade partners, accounting for 0.8\% of total exports.} Therefore, by looking from the perspective of a single exporting country, France, we would expect a non-negligible impact, possibly magnified by the competition of French exporters with other European producers.

We proceed with our investigation in three steps. At first, we evaluate the overall impact of CETA at the product level. Crucially, at this stage, we find that CETA positively impacted French exports at both the intensive and extensive margins. On the one hand, product-level flows to Canada increased by an average of 6.4\% on a sales-weighted basis. On the other hand, we find that there has been significant product churning beyond regular entry-exit dynamics due to the free trade agreement, as about 7.3\% of new French products reached Canada for the first time and 8.1\% of them abandoned the market due to the new provisions.

Notably, our matrix completion approach allows us to expose all the relevant heterogeneity after the impact of a trade treaty. In fact, we observe that the CETA can have either positive or negative impacts on each treated unit. This means we can evaluate products (or firms) that incur either gains or losses as a result of the trade agreement. We argue that this is an advantage over other estimators, which usually summarize the impact after evaluating only one parameter, or just a few when they control for heterogeneous effects. In our case, we have a full distribution of idiosyncratic treatment effects, one for each element of the trade matrix. 

After a post-estimation analysis, we observe that product-level idiosyncratic treatment effects are positively associated with a measure of revealed comparative advantage of French exporters vs. the rest of the world. That is, the increase in export flows has been higher for those products for which French producers had a competitive edge before the treaty's signature, and they have reduced the export flows for products at a comparative disadvantage. This aligns with standard trade theory. According to the \citet{bernard2007comparative} framework, trade liberalization triggers industry shrinkage and net job losses in sectors of comparative disadvantage. Because these sectors see a smaller share of firms survive as exporters, productivity gains are less significant than in advantaged industries. Ultimately, this domestic resource reallocation shifts the economy's focus, further reinforcing its primary comparative advantage.

The remainder of the paper is structured as follows. We begin with a short review of the relevant literature in Section \ref{sec: literature}. Section \ref{sec: data} presents the data and offers preliminary evidence. In Section \ref{sec: methods}, we outline the empirical strategy. Results are displayed in Section \ref{sec: analysis}, while robustness and sensitivity checks are presented in Section \ref{sec: robustness}. Section \ref{sec: conclusion} concludes.

\section{Related Literature}
\label{sec: literature}

\textit{Ex-post} evaluation of free trade agreements is challenging \citep{Baier_et_al_2019} because they often entail an endogenous selection of partners or products \citep{BAIER200429,baier2009estimating}, on the one hand, and a self-selection of heterogeneous exporters \citep{melitz2003impact}, on the other hand. Hence, \citet{goldberg_2016} consider this endogeneity a major hurdle to the causal identification of the economic impact of PTAs.

This endogeneity of PTAs has been addressed by using various approaches, including gravity equations with additional controls (e.g. bilateral fixed effects) for unobserved characteristics \citep{aitken1973effect,abrams1980international,bergstrand1985gravity,soloaga2001regionalism,feenstra2001using}, instrumental variable (IV) or control-function techniques with cross-sectional data \citep{baier2002endogeneity, magee2003endogenous, baier2009estimating}, panel data models with a rich set of fixed effects \citep{head1998immigration,baier2007free, westerlund2011estimating, YANG2014138}, or matching techniques \citep{baier2009estimating, Egger_Tarlea_2020}.\footnote{See the reviews by \citet{limao_2016} and \citet{Larch_Yotov_2023} of the empirical exercises estimating the impact of trade agreements.}

In this paper, we explore the scope for using a potential outcome model to assess the causal impact of preferential trade agreements.\footnote{The framework for causal inference that uses ‘potential outcomes’ to define causal effects at the unit level in the context of randomized experiments and quasi-experiments is dubbed the Rubin Causal Model \citep{Rubin2005}. The introduction of this framework in economics helped comply with the so-called \textit{credibility revolution} cited by \citet{Angrist_Pischke_2010}.} In particular, we draw from the most recent advances in causal machine learning, whose aim is to estimate average causal effects after predicting the missing potential outcomes with non-parametric methods \citep{abadie2010synthetic, abadie2015comparative, arkhangelsky2019synthetic, chernozhukov2021exact}. Specifically, we leverage the literature on matrix completion, which originally exploited observed information to predict unobserved entries when matrices are sparse  \citep{candes2010matrix,candes2012exact,mazumder2010spectral}. For our purpose, we adapt the algorithm proposed by \citet{athey2021matrix}, whose intuition is that a matrix approach can also be used for causal inference while allowing for an exogeneity assumption that is weaker than the one in methods using control groups \citep{Poulos_et_al_2021, Xu_2017, Athey_2015}.

On top of empirical challenges, we know from trade theory that opposing mechanisms may hinder an accurate estimate of the impact of tariff reduction. On the one hand, a tariff reduction implies greater market access because the demand increases in the liberalized market. On the other hand, tariff reductions under trade agreements may have pro-competitive effects. When Marshall's second law of demand does not apply, monopolistic exporters may reduce their markups in response to reduced tariffs \citep{mrazova_2017} or preferential market access \citep{Crowley_Han_2022}. This induces, in turn, selection effects. Market size and trade openness affect the intensity of competition in a market, which reinforces the selection of exporters to that market \citep{melitz_ottaviano_2008}.  Against this background, we design our empirical strategy to encompass multidimensional counterfactuals at both the product and firm levels, which enable us to discuss competing mechanisms.

Crucially, our empirical design acknowledges the role of heterogeneous firms in trade agreements, especially in a world where multi-product firms dominate trade flows \citep{feenstra2007optimal, eckel2010multi, iacovone2010multi,bas2013chinese}. 
In this, we refer to
\citet{mayer2014market} and \citet{bernard2010multiple,bernard2011multiproduct}, who incorporate multi-product firms into models of heterogeneous firms while building upon the pioneering work by \citet{melitz2003impact}. They show that tougher competition in a liberalized market leads firms to skew their export sales towards their better-performing products. On a similar line of research, \citet{dhingra2013trading} and \citet{qiu2013multiproduct} predict that falling trade costs make the most productive firms expand their product scope, while the least productive firms contract theirs. According to \citet{baldwin2009impact}, the net effect could be ambiguous because tariff cuts can both increase exporters' plant size by extending the production run length of the exported portion of the product line and reduce exporters' plant size by reducing the total number of products.

\section{Data and preliminary evidence}\label{sec: data}
\subsection{Customs data and trade regime changes}

Our primary data source is the French Customs (\textit{Direction Générale des Douanes et Droits Indirects})\footnote{We accessed the database through the CASD, French Secure Data Access Center (project DYNAMEX).}, which contains records of export sales for each product-firm-destination-month observation. Products are originally classified by the 8-digit Combined Nomenclature (CN8), and firms are identified by their \textit{SIREN} number, i.e., the 9-digit identifier assigned to every registered business in France by the National Institute of Statistics and Economic Studies. Moreover, we rely on the WTO tariff databases to retrieve information on those products at the HS 6-digit level whose tariffs or quotas have been modified by the EU-Canada Comprehensive Economic and Trade Agreement (CETA).\footnote{Appendix Table \ref{tab:tariff changes} briefly summarizes the extent of tariff changes for French exporters in Canada due to CETA.}

Original customs data are first aggregated from monthly to yearly levels in the September-August segments, following the timeline of the trade treaty, which became operational in September 2017. In addition, we align the product classification from the 8-digit Combined Nomenclature (CN) to the 6-digit Harmonized System (HS) classification to match the original information on products whose tariff or tariff quota has been changed by CETA. Since the HS classification was revised in 2017, we have converted back to HS 2012. 

So far, we have identified the perimeter of the product-level analyses we perform in Section \ref{sec: product_level}. In the first part of our paper, our investigation encompasses all products that French exporters direct to Canada, regardless of the firms' characteristics. In the second part of the empirical strategy, we will focus on multiproduct firms; therefore, we need to eliminate from our sample perimeter\footnote{In the original data, we find firms that are active in service industries and occasionally export goods. We eliminate these cases from our firm-level sample perimeter because they conceal a delivery of materials needed to proceed with the service supply (e.g., building materials for construction firms, laboratory equipment for an R\&D company, etc.) At the same time, in the firm-level exercise, we keep trade intermediaries, who traditionally export on behalf of other producers, and we compare results with manufacturing exporters.}: i) firms that do not export to Canada, ii) firms that export only one product to Canada.

\begin{figure}[ht]
    \centering
    \caption{Products' coverage in 2016}
     \resizebox{0.55\textwidth}{!}{%
\includegraphics[width=\textwidth]{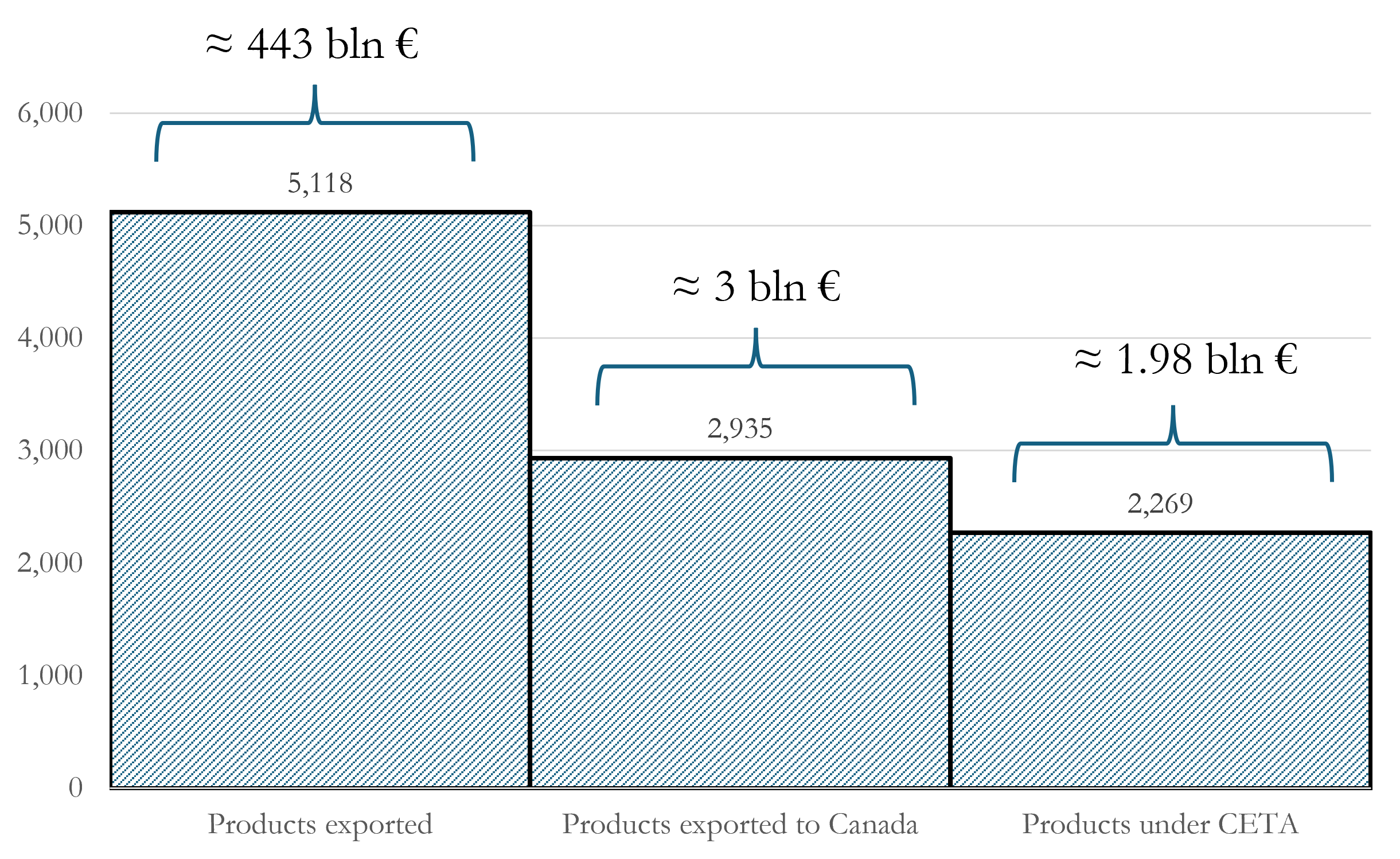}}
   \label{fig:coverage products}
   \begin{tablenotes}
   \footnotesize \singlespacing
\item Note: The figure shows sample coverage of products in 2016. The y-axis indicates the number of products, whereas the text boxes on top of the bars indicate the total trade value in 2016. On the left is the number of products exported from France to any destination. In the centre is the number of products exported to Canada. On the right is the number of products that are both exported to Canada and fall under the provisions of the Canada-EU Trade Agreement.
\end{tablenotes}
\end{figure}

\begin{figure}[ht!]
    \centering
    \caption{Firms' coverage in 2016}
     \resizebox{0.55\textwidth}{!}{%
\includegraphics[width=\textwidth]{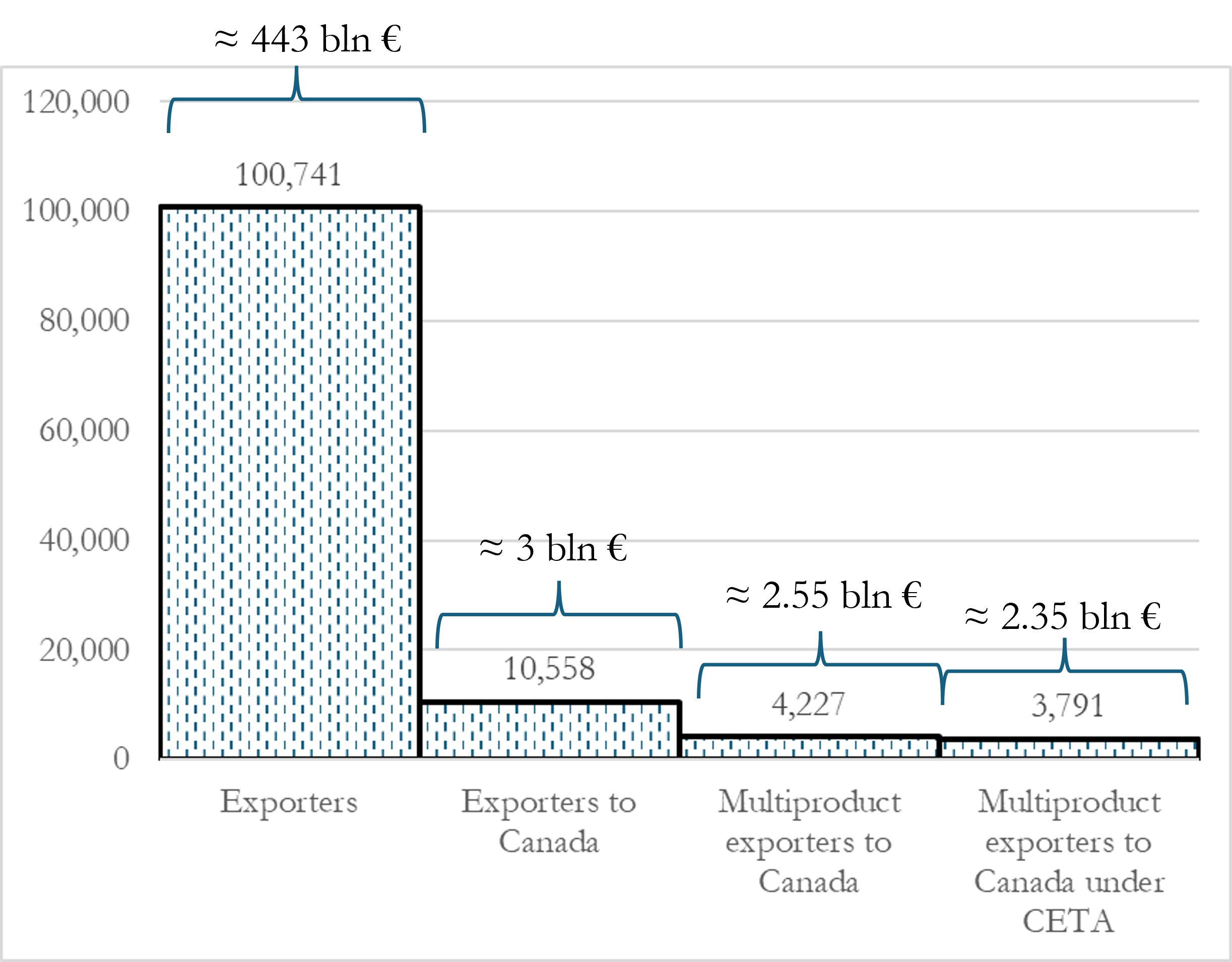}}
   \label{fig:coverage firms}
      \begin{tablenotes}
      \footnotesize \singlespacing
\item Note: The figure shows sample coverage of exporters in 2016, while text boxes on top of the bars indicate the total trade value in 2016. On the left is the number of French firms that exported to any destination. Then, we report the number of exporters to Canada and, among them, the number of multiproduct firms that export at least two products to Canada. On the right is the number of multiproduct exporters to Canada, with at least one product listed by the Canada-EU Trade Agreement, for which we indicate the value of their total exports to Canada, encompassing both products with and without a trade regime change.
\end{tablenotes}
\end{figure}

In Figures \ref{fig:coverage products} and \ref{fig:coverage firms}, we present waterfall charts to visualize the relevance of the products and firms included in our study relative to 2016, just before the CETA. On the one hand, when we separate products liberalized after CETA, we observe that they make up 77\% of the total product lines exported from France to Canada. On the other hand, the list of products that have seen a change in the tariff or non-tariff regime thanks to CETA coincides for about 57\% with the list of product lines that French exporters already trade with the rest of the world. 
From our perspective, either stylized fact is worth further investigation. In the first case, we expect endogenous product selection during treaty negotiations, and we test this in the following paragraphs.

As for firms, we first need to exclude those that have never exported to Canada, as they are not directly affected by the CETA signature. Then, following a basic definition of multiproduct firms, we will consider only those that export at least two products to Canada. In this case, as shown in Figure \ref{fig:coverage firms}, only about 10.5\% of French exporters reach Canada as an export destination. Among them, about 40\% are multiproduct firms with a portfolio of at least two products in Canada. Finally, among the latter, 79.8\% have seen a tariff or non-tariff change in at least one of their products exported to Canada after CETA.

In the second part of the paper, the subset of multiproduct firms is of special interest to us not only because they are relevant in terms of aggregate trade flows (2.55 billion euros vs 3 billion euros of total exports to Canada) but also because they are a segment that potentially shows adjustments in product scope, which would be otherwise hidden if we do not consider the firm-level dimension. In Appendix Figure \ref{fig:products per exporter}, we show French exporters' distribution of product portfolios to Canada.

\subsection{Preliminary evidence}
\label{sec: preliminary_evidence}
In the following paragraphs, we investigate whether products and firms that have experienced a change in the trade regime differ significantly from those that have not. The obvious intuition is that negotiators could have chosen production segments that offer higher gains from trade. Alternatively, larger firms may have had the power to impose their agenda on negotiators. In Table \ref{tab: endogenous selection}, we investigate the issue with two sequences of t-tests on the difference in the means after a choice of a set of indicators that could possibly capture the peculiar differences between products included and not included in the CETA. First, we test bilateral exports from France to Canada. Then, we consider the same product partition under the CETA, this time examining the features of products and producers at the global level after aggregating across destinations.

\begin{table}[!ht]
\caption{Characteristics of trade flows before CETA - 2015M01-2016M12}
    \centering
       \resizebox{.7\textwidth}{!}{
    \begin{tabular}{lccc}
    \hline
        & products & products & difference\\
        & in the CETA & not in the CETA & in means \\ \hline \hline
                  ~ & ~ & ~ & ~ \\
        \textit{Exports to Canada} & ~ & ~ & ~ \\
                ~ & ~ & ~ & ~ \\
        Avg. trade value & 30231.8 & 54023.6 & -35700.5*** \\
        Avg. dispersion & 65579.8 & 122571.7 & -78671.4*** \\ 
        Avg. number of transactions & 2571.4 & 599.9 & 1971.4*** \\ 
        Avg. number of firms & 212.1 & 100.2 & 111.8*** \\
        Avg. firm's exports & 509,037.5 & 207,466.9 & 301,507.6*** \\
        ~ & ~ & ~ & ~ \\
        \textit{All exports} & ~ & ~ & ~ \\
        ~ & ~ & ~ & ~ \\
        Avg. trade value & 35265.7 & 60645.2 & -25379.5*** \\
        Avg. dispersion & 162147.3 & 301385.0 & -188687.6*** \\
        Avg. number of transactions & 42852.1 & 23216.9 & 19635.2*** \\
        Avg. number of firms & 1290.5 & 1278.3 & 12.18*** \\
        Avg. firm's exports & 8,150,142 & 1,412,479 & 6,737,762*** \\        
\hline
    \end{tabular}}\label{tab: endogenous selection}
    \begin{tablenotes}
    \footnotesize
        \item Note: The table reports t-tests computed on average indicators of the export matrix in 2015-2016, considering products that will see a change with CETA in 2017 (column 2) vs. products whose trade regime will not change (column 3). Column 4 reports differences in the means considering unequal variances. *** stands for $p \leq 0.001$, hence the average means are significantly different. In the first half of the table, we consider only export flows to Canada, i.e., the destination involved in the treaty. In the second bottom half of the table, we expand the matrix to include export flows to all export destinations, although they are not parties to CETA.
    \end{tablenotes}
\end{table}

The first three indicators we test in Table \ref{tab: endogenous selection} refer to features of the product-level monthly flows observed in the period 2015M01 to 2016M12, while the other two indicators refer to the firm-level dimension. Starting from the top of the table, we observe that the average trade value of products included in the CETA was lower, with less dispersion around the sample means, and that transactions were more frequent in the two years preceding the treaty's signature. Among exporters, the product was traded by more firms that, on average, had a relatively higher exposure to Canada as an export destination.

If we look at the bottom of the table, we see that the same differences observed in the bilateral relationship between France and Canada are confirmed by aggregate flows between France and the rest of the world. Briefly, the products included in the CETA are usually traded by firms whose export size is, on average, larger, while single monthly flows are smaller, more frequent, and less volatile around the mean in the two years before the CETA.

Eventually, preliminary evidence shown in Figure \ref{tab: endogenous selection} motivates the choice of an empirical strategy that is capable of handling an endogenous selection of product lines in a trade treaty, thus making policy evaluation unbiased by the political economy of larger firms or by the tendency of negotiators to cherry-pick products that already have higher potential.

\begin{figure}[htpb]
     \centering
      \caption{Time trends at the product level, intensive and extensive margins}
     \begin{subfigure}{0.40\textwidth}
         \centering
         \includegraphics[width=\textwidth]{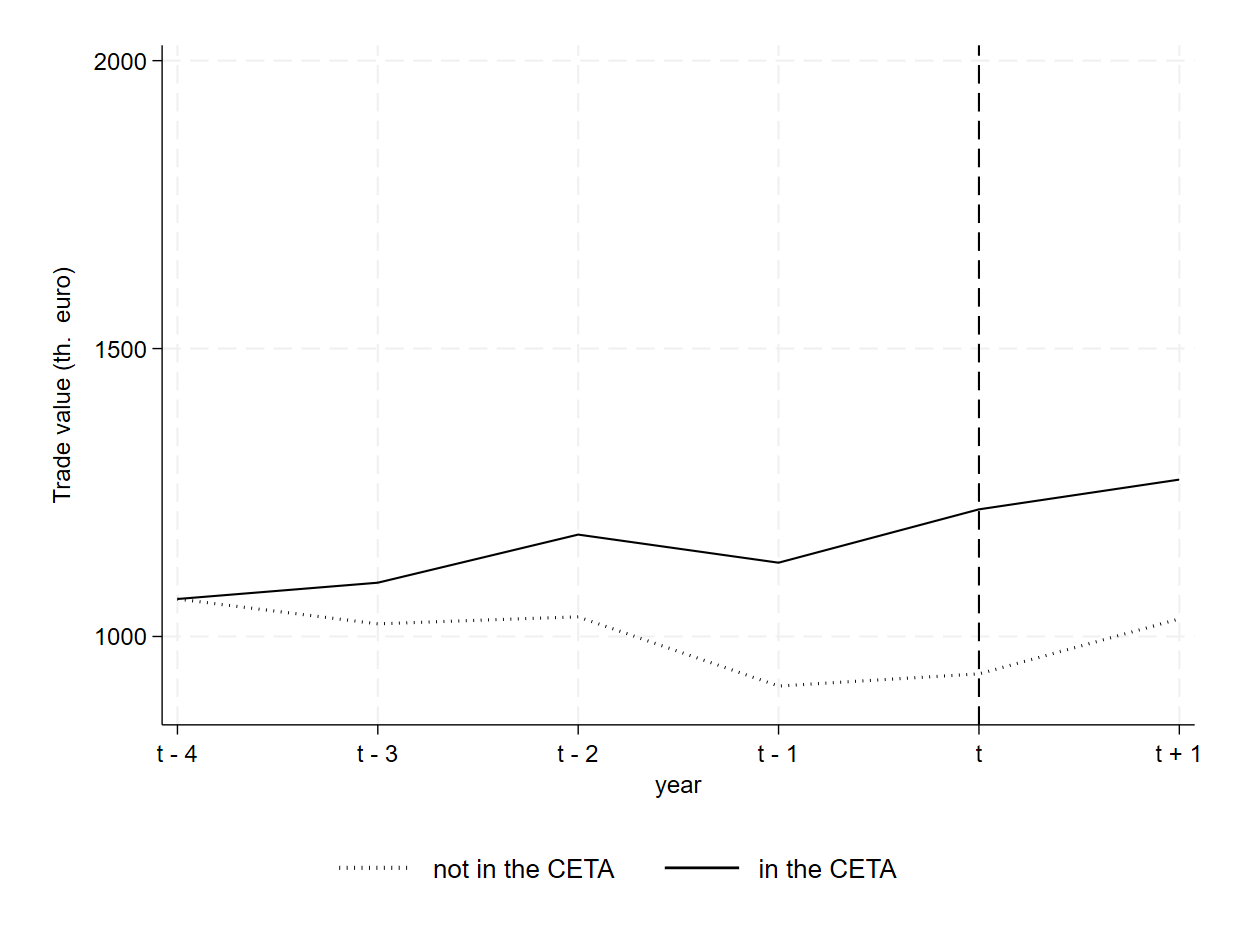}
         \caption{Intensive margin }
     \end{subfigure}
     \begin{subfigure}{0.40\textwidth}
         \centering
         \includegraphics[width=\textwidth]{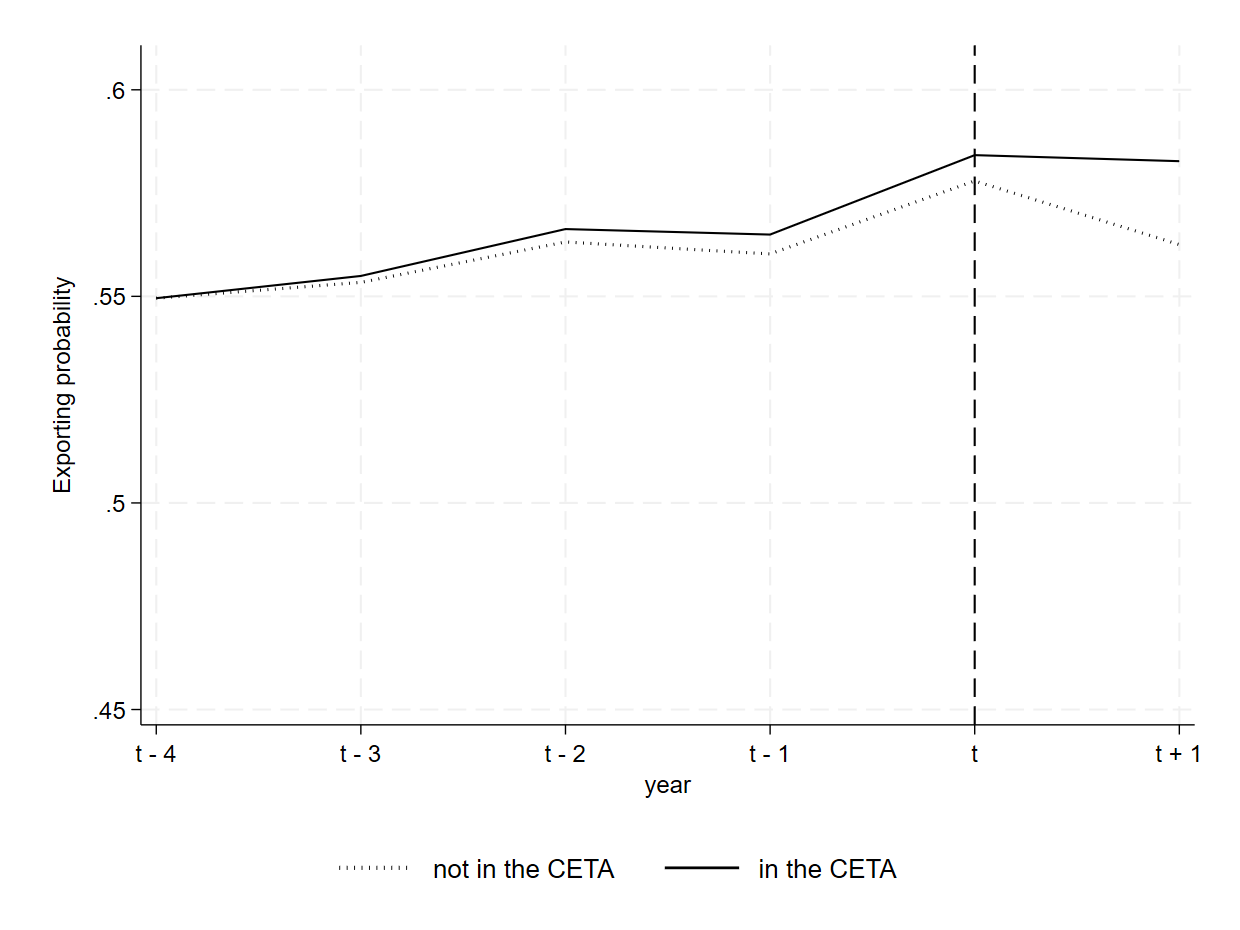}
         \caption{Extensive margin}
     \end{subfigure}
    \begin{tablenotes}
        \singlespacing
        \footnotesize
        \item Note: We report in panel (a) linear trends for exports of product lines to Canada, separating those that are included in the CETA and those that are not. In panel (b), we report linear trends in the probability that a new product line is exported to Canada, separating those included in the CETA from those not included. The graphs are generated using the predictions of a difference-in-differences model augmented with interactions between time and an indicator for treatment when products are enlisted under the CETA.
    \end{tablenotes}
        \label{fig:parallel_trends_products}
\end{figure}
\begin{figure}[htpb]
     \centering
      \caption{Time trends at the firm-level on the intensive and extensive margins}
        \begin{subfigure}{0.40\textwidth}
         \centering
         \includegraphics[width=0.99\textwidth]{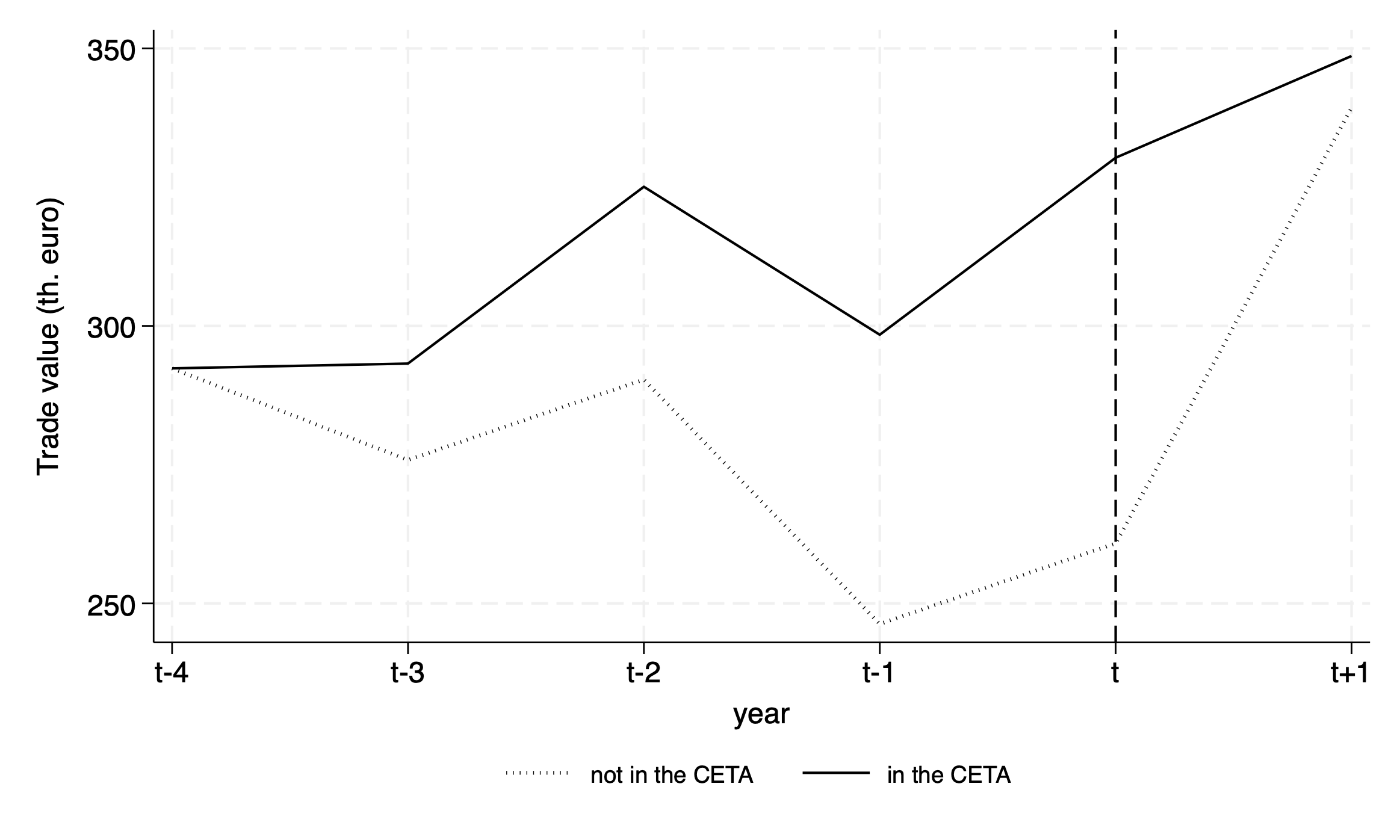}
         \caption{Intensive margin}
     \end{subfigure}
     \begin{subfigure}{0.40\textwidth}
         \centering
         \includegraphics[width=\textwidth]{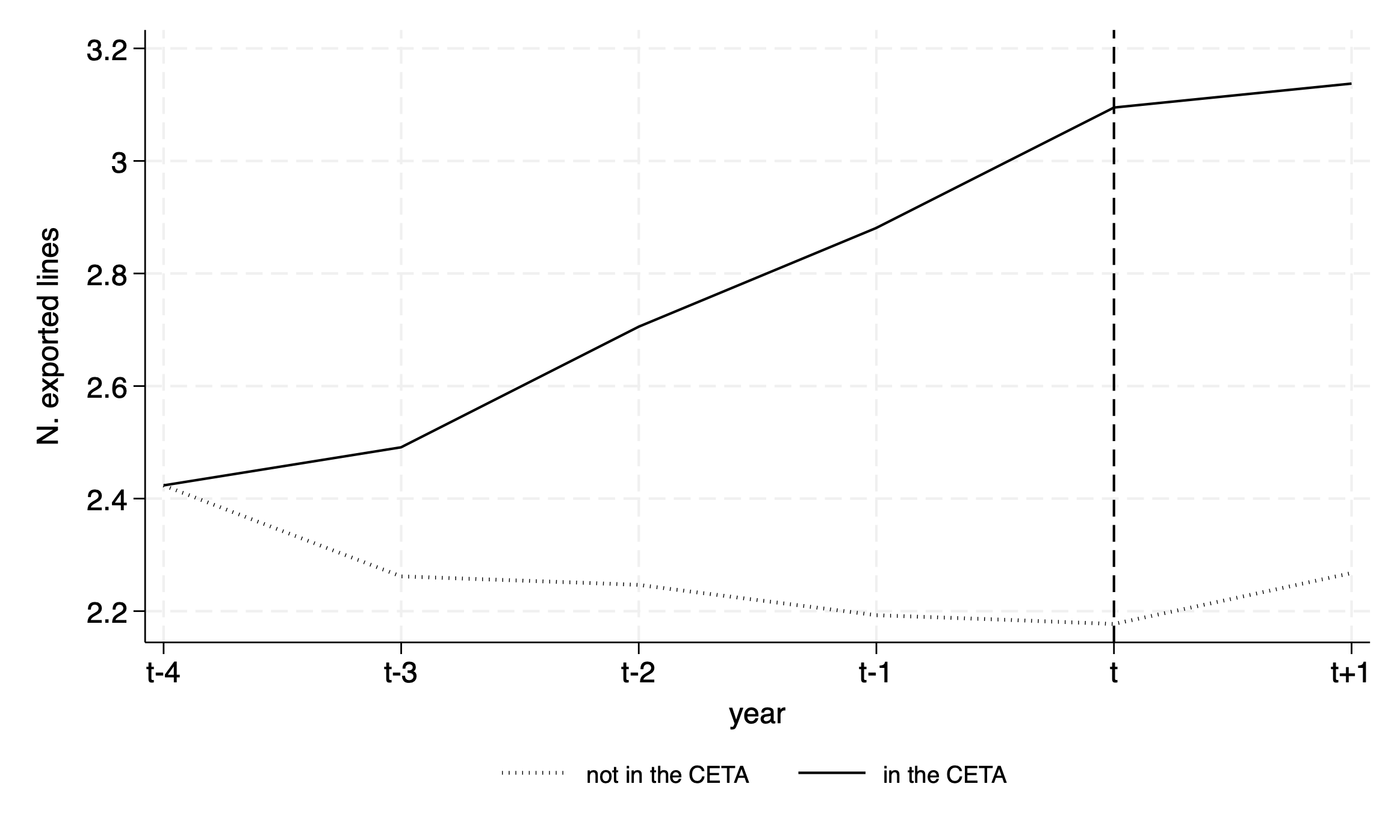}
         \caption{Extensive margin}
         \label{fig:parallel_ext_CA_firm}
     \end{subfigure}
    \begin{tablenotes}
        \singlespacing
        \footnotesize
        \item Note: We report in panel (a) linear trends for exports of firms to Canada, separating those that have a product enlisted by the CETA and those that have not. In panel (b), we report linear trends in the number of lines a firm exports to Canada, separating those with a product included in the CETA from those without. The graphs are generated using the predictions of a difference-in-differences model augmented with interactions between time and an indicator for treatment when firms have a product listed by the CETA.
    \end{tablenotes}
        \label{fig:parallel_trends_firms}
\end{figure}

Our preferred empirical strategy should also be able to handle heterogeneous time trends. It is, in fact, possible that the products and firms affected by the CETA were already on a path to growth before the treaty was signed. The presence of unparallel time trends could possibly confound the actual impact of the trade treaty. In Figures \ref{fig:parallel_trends_products} and \ref{fig:parallel_trends_firms}, we display linear trends after the estimation of simple difference-in-difference models\footnote{We estimate simple difference-in-difference models augmented with terms that capture the differences in slopes across the products/firms that are concerned by the CETA and those that are not. See Appendix B for more details. Results of the difference-in-difference models are reported in Appendix Table \ref{tab:diff_in_diff}. Please note that the diff-in-diff results suggest that the CETA had only an effect on the firm-level extensive margin, whereas no significant impact is observed on the intensive margins at the product and firm levels. While serving as a valuable reference point, a simple diff-in-diff methodology cannot be valid if the assumption of parallel trends is violated, as from Figures \ref{fig:parallel_trends_products} and \ref{fig:parallel_trends_firms}, and when the treatment is not orthogonal to relevant characteristics of the treated units, as from Table \ref{tab: endogenous selection}.} of the intensive and extensive margins for both products and firms, separating when they are concerned by the CETA and when they are not.

After a graphical inspection, we can observe that intensive margins at the product and firm levels (panels (a) in Figures \ref{fig:parallel_trends_products} and \ref{fig:parallel_trends_firms}) were already on diverging paths. For products, those not included in the CETA were already in a downward trend. In the case of firms, those that did not have a product listed in the CETA had been in a declining trend in the years before the treaty, and then increased significantly thereafter. In the case of extensive margins, product flows show no significant differences, whereas firm-level pre-trends were significantly divergent.

In Section \ref{sec: exogeneity}, we will discuss the conditions under which our matrix completion estimates can handle both the endogenous selection into treatment and the lack of parallel trends that have been detected in the previous paragraphs. From the previous perspective, it is already clear how the traditional differences-in-differences method can yield biased results. Similarly, any quasi-experimental approach that relies on a simple comparison between treated and untreated observations could be biased.

\section{Empirical strategy}
\label{sec: methods}

\subsection{Treated products and treated firms}
In the following paragraphs, we develop an empirical strategy to evaluate the impact of CETA. For the sake of generalization, we will define a generic $u$-th unit of observation at time $t$, such that the exposure to CETA, i.e., our treatment, can be defined as $W_{ut}$. It allows us to introduce two different definitions of a policy treatment: at the product-destination level and at the firm level.

At the product-destination level, we will consider the treated population, $\mathcal{T}$, consisting of all the products that experienced a tariff or a quota change after CETA. Let $p$ denote the product, $d$ represent the destination, and $t$ indicate time. Notice that $d$ can indicate either Canada, as the destination where we actually observe a tariff or quota change, or an alternative destination where the same products are exported from France. Please note that we consider a product as treated regardless of the destinations in which it is exported. The latter setup will prove useful to evaluate trade diversion/creation effects later in the paper. 

Since CETA entered into force in September 2017, we aggregate monthly flows by year $\tau$ in the period September-August.\footnote{In the following, $\tau-5$ refers to the period from September 2012 to August 2013, $\tau-4$ refers to the period from September 2013 to August 2014, and so on, unti the treatment period, $\tau$, which refers to the period from September 2017 to August 2018. Our dataset covers information up to December 2018, so we can only observe one period ($\tau$) ahead of CETA. Consequently, the analysis is restricted to the Treaty's short-term effects. Nonetheless, our approach is also suitable for analyzing a staggered adoption scheme across multiple post-treatment periods.} In this case, the treatment indicator is defined as follows:
\begin{equation}\label{eq: product treatment}
    W_{pdt}=\begin{cases}
        1& p \in \mathcal{T},t\geq\tau\\
        0& otherwise
    \end{cases}
\end{equation}

When we switch to the firm level, our population consists of multi-product firms that export at least two distinct products to Canada.\footnote{See Section \ref{sec: data} for a description of the firm-level sample selection strategy.} Among them, the set of treated firms $\Theta$ is defined as: 
\begin{equation*} \label{eq: treated firms}
    \Theta=\{i :\ \Psi_{itCA}\cap \mathcal{T}\neq \emptyset, t=[\tau-5,\tau],\}
\end{equation*}

where $\Psi_{itCA}$ represents the set of products $p$ exported to Canada by firm $i$ in year $t$, and $|\Psi_{itCA}| \ge 5$. Briefly, we consider any firm treated if it exported at least two products to Canada, at least one of which is listed in CETA, before or after the entry into force of CETA. Conversely, we will consider non-treated those firms that exported at least two products to Canada before CETA but do not have any products included in their portfolio in the CETA. 

Once we have defined the set of \textit{treated firms}, $\Theta$, we can establish the treatment at the firm-\textit{per}-product level. Let $i$ denote the firm, $p$ indicate the product, and $t$ represent the year. The treatment indicator at the firm level is defined as:
\begin{equation} \label{eq: firm treatment}
       W_{ipt}=\begin{cases}
       1 & \forall\ |\Psi_{itCA}| \ge 5,\ i\in \Theta,\ t\geq \tau\\
       0& otherwise
       \end{cases}
\end{equation}.\\

Therefore, in the following paragraphs, when we deem it unnecessary to specify, our generic indicator of treatment $W_{ut}$ for the $u$-th unit will suffice. When presenting results, we will indicate which of the eqs. \ref{eq: product treatment} or \ref{eq: firm treatment} defines the treatment.

\subsection{Matrix completion}
\label{sec: matrix_completion}

At this point, we are ready to illustrate the details of our application on trade policy evaluation. Originally, matrix completion methods were used to recover lost information in highly sparse matrices. In the context of statistical and computational science exercises, the task has been to fill in the missing entries of a partially observed matrix \citep{candes2010matrix,mazumder2010spectral,candes2012exact}. The novel intuition by \citet{athey2021matrix} is that one could instead frame a matrix completion algorithm in the context of potential outcome models, where predictions can be obtained for missing multidimensional counterfactuals. Here we adapt the framework by \citet{athey2021matrix} to the case of a trade policy, where we have $N$ units of observations, $T$ time periods, and there exists a pair of potential outcomes, $Y_{ut}(0)$ and $Y_{ut}(1)$, with unit $u$ exposed in period $t$ to the entry into force of the CETA. The generic treatment has been defined in the previous section as a matrix with entries $W_{ut} \in \{0, 1\}$, and the realized outcomes are thus equal to $Y_{ut} = Y_{ut}(W_{ut})$.

In our case, the fundamental problem of causal inference is that a set $\mathcal M < NT$ of potential outcomes is not observed. Specifically, we do not observe the outcomes of the treated units as if the treatment did not occur. This means that we will never observe the potential exports of products/firms concerned by the CETA as if the latter had not been signed. Yet, we can obtain valid counterfactuals for the set $\mathcal M$, and the solution is to predict them using the information available in the trade matrix from entries $\mathcal O \equiv NT - \mathcal M$, which are observed. Once we obtain valid counterfactuals, we can compute the idiosyncratic\footnote{Please note that, at this stage, treatment effects are numerical values, not random variables anymore. They have specific internal validity for the exercise, but external validity is not guaranteed.} Treatment Effect on the Treated (TET) is expressed in monetary values as:

\begin{equation}\label{eq: tet}
\forall   \{u, t\} \in \mathcal M: TET_{ut}= Y_{ut}(1)- \hat{Y}_{ut}(1) 
\end{equation}.

Then, we can manipulate the latter expression to find the best solution, in levels or in percentage points. In the next paragraphs, it will depend on whether we want to comment on the intensive or the extensive margin, as explained below.  

\subsubsection{Effects on the intensive margin}

We can evaluate the impact of the new trade regime on the intensive margin by examining the moments of the entire distribution generated by the entries in the counterfactual matrix. In this case, we prefer to express the idiosyncratic treatment effect on treated from eq. \ref{eq: tet} as a ratio, to comment in relative terms and on percentage points, in the form: 
\begin{equation}
\forall   \{u, t\} \in \mathcal M: TET_{ut}^*=\frac{Y_{ut}(1) - \hat{Y}_{ut}(1) }{Y_{u,t-1}(1)} \times 100  
\end{equation}

where $Y_{ut}$ is the observed value for unit $u$ at time $t$, $\hat{Y}_{ut}$ is corresponding predicted value, and $Y_{u,t-1}$ is the observed value for unit $u$ at time $t-1$.

Finally, we can compute the weighted average treatment effect on the treated (wATET), also expressed in relative terms, in the form:

\begin{equation}\label{eq: wATET}
wATET =  \sum_{\{u,t\} \in \mathcal{M}} s_{ut} TET_{ut}^*
\end{equation},

where $s_{ut}$ indicates the salience of the export flows. For the sake of simplicity, we can use for salience the share of the trade flows of unit $u$ at time $t-1$, i.e., before the signature of the CETA, on the total export flows for each entry $\{u,t\} \in \mathcal M$.

\subsubsection{Effects on the extensive margin}

In the evaluation of the trade extensive margin, the potential outcomes are binary: $Y_{ut}(1) = \{0,1\}$, i.e., they are equal to 1 if the product is exported and 0 otherwise. Our matrix completion application is reduced to a classification problem, and we obtain predictions in a binary form, $\hat Y_{ut}(1) = \{0,1\}$, such that idiosyncratic Treatment Effects on the Treated can have three alternative values, $TET_{ut} \in \{-1, 0, 1\}$. A value $-1$ means that our counterfactual predicts that a trade flow existed in that entry of the trade matrix, but it actually did not. We will define the latter as the negative extensive margin. A value of $1$ indicates that our counterfactual holds: the product should not have been traded, but it actually was. We will call the latter a positive extensive margin. On the other hand, whenever we find $TET_{ut}=0$, it means that our predictions and the observed outcomes correspond. Please note that, against the previous background, products can still enter or exit the foreign market following regular product churning, regardless of a change in the trade regime. The latter cases would all be flagged with a zero in the set of idiosyncratic treatment effects.

\subsubsection{The estimator}
Let us start by representing the entire trade matrix. In the product-level analysis, we will have a matrix with entries defined by the trade value of each 6-digit product-\textit{per}-destination (i.e., the $u$-th observation) and time in a cell. In the firm-level analysis, we report each matrix cell's trade by firm-\textit{per}-product (i.e., the $u$-th observation) and time. Next, we empty the set $\mathcal M$ of matrix entries where we have exports with tariff and tariff-quota changes after the CETA signature, i.e., ${Y}_{ut}(1)$ when $\geq 2017$, and we ask the algorithm to reconstruct the full matrix while feeding it information from the set $\mathcal O$, including:

\begin{enumerate}
    \item treated and untreated observations before the treatment, when CETA did not exist (i.e., $Y_{ut}(1)$ and ${Y}_{ut}(0)$ when $t<2017$)
    \item untreated observations after the treatment (i.e., ${Y}_{ut}(0)$ when $\geq 2017$)
\end{enumerate}

Further details on the product-level and firm-level trade matrices are described in Sections \ref{sec: product_level} and \ref{sec: firm_level}, respectively. In our context, the value of a matrix completion approach lies in its ability to leverage all available information nonparametrically without making stringent assumptions about joint distributions, parallel trends, and functional forms. Notably, we still rely on the classical SUTVA assumption (see Section \ref{sec: exogeneity}), according to which our treatment is only binary and there is no interference, i.e., treated units do not spill over on untreated units, conditional on the information derived from the model. These conditions hold for each quasi-experimental approach in which we assume we can evaluate treated vs. untreated units.

By predicting each unobserved potential outcome, we obtain multidimensional counterfactuals for each cell in a matrix pertaining to treated units, thereby taking on board all the heterogeneity that can possibly derive from a trade policy treatment.\\

We obtain predictions from a decomposition of the $N\times T$ matrix $\mathbf{Y}$, such that:
    \begin{equation}
    \mathbf{Y} =\mathbf{\tilde Y} +\mathbf{\tilde\gamma}+\mathbf{\tilde \delta}+\mathbf{\varepsilon} 
    \label{eq: model}
\end{equation}
where we can collect $\mathbf{\hat Y}=\mathbf{\tilde Y} +\mathbf{\tilde\gamma}+\mathbf{\tilde \delta}$, as these are the components we want to estimate. Among them, $\mathbf{\tilde Y}$ is a low-rank matrix with respect to the original $N\times T$. Then, we have $\mathbf{\tilde\gamma}$, which is the $N\times1$ vector of row-fixed effects, and $\mathbf{\tilde \delta}$, which is the $1\times T$ vector of time fixed effects.\footnote{Note that the row and column-fixed effects can be subsumed in matrix $\mathbf{\tilde Y}$. However, \citet{athey2021matrix} already noted that separating fixed effects without regularization greatly improves predictive performance. In our case, we confirm that prediction power deteriorates when we do not separate fixed effects.} In our context, the $N\times1$ vector of row-fixed effects can represent either product-destination or firm-level fixed effects, respectively. It depends on which exercise we consider. $\varepsilon$ is an $N\times T$ matrix of random noise values.

Our $\mathbf{\tilde Y}$ is the result of a singular value decomposition (SVD), such that $\mathbf{\tilde Y}=\mathbf{S} \mathbf{\Sigma} \mathbf{R^\top}$, where $\mathbf{S}$ and $\mathbf{R}$ are unitary matrices, and $\Sigma$ is a rectangular diagonal matrix with singular value entries $\sigma_u(Y)$. The latter entries are substituted by $\max(\sigma_{i}(\mathbf{\tilde Y}) - \lambda_Y, 0)$ after regularization.  In fact, we introduce regularization on the $\mathbf{\tilde Y}$ component, $\lambda_Y \vert\vert\mathbf{\mathbf{\tilde Y}}\vert\vert$, to avoid overfitting. In our context, overfitting implies that the model fits the training data too closely and its predictive power will be poor for counterfactuals. Indeed, overfitting problems are more likely to arise in cases like ours, where we have a high $N \times T$ dimensionality. Finally, the estimator can be written as the result of an optimization problem in the general form:

    \begin{equation}\label{eq: generic}
        \min_{\tilde Y,\gamma, \delta}\Bigg[\sum_{(u,t)\in \mathcal{O}} \frac{1}{|\mathcal{O}|}\Bigg(Y_{ut}-\tilde Y_{ut} -\gamma_u-\delta_t\Bigg)^2+\lambda_Y||\mathbf{\mathbf{\tilde Y}}||_*\Bigg]
        \end{equation}

where $\mathcal{O}$ includes any pair $(u, t)$ in the set of observed export outcomes, and $\vert\vert \mathbf{\tilde Y} \vert\vert_*$ is the nuclear norm of the matrix $\Tilde{\Sigma}$ resulting from shrinking the scaling matrix with the singular value decomposition (SVD) by $\lambda_Y$. We select the optimal value of $\lambda_Y$ after cross-validation\footnote{As we choose a nuclear norm for regularization, the estimator can be computed using fast convex optimization programs like the one proposed by \cite{mazumder2010spectral}.} on $K$ different random subsets $\mathcal{O}_k \subset \mathcal{O}$ of the original matrix, having a fraction of observed data equal to the one in the original sample. Finally, once we have predicted matrix $\mathbf{\hat Y}$, we obtain the counterfactuals we need to estimate treatment effects as in eq. \ref{eq: tet}.

\subsubsection{The exogeneity assumption}\label{sec: exogeneity}
In the previous paragraphs, we introduced an application of matrix completion by \citet{athey2021matrix} to trade data. In the following paragraphs, we will assess its predictive power by reproducing the observed treated observations. Prediction power is useful for gauging how well the method predicts counterfactuals. Yet, a high prediction power alone does not make the estimator reliable for causal interpretations. In this section, we define the exogeneity assumption under which we can interpret the treatment's impact on individual observations. We have the following equation\footnote{A general case on semi-parametric difference-in-difference estimators in the presence of fixed effects was already made by \citet{Abadie_2005}.}:

\begin{equation}\label{eq: exogeneity}
\E[\epsilon_{ut} | \tilde Y_{ut}, \gamma_{u}, \delta_{t}]=\E[\epsilon_{ut} | W_{ut}, \tilde Y_{ut}, \gamma_{u}, \delta_{t}]=0
\end{equation}

where the expected value of errors, $\epsilon_{ut}$, is zero conditional on the low-rank matrix to be estimated, $\tilde Y_{ut}$, the fixed effects on individual units, in our analyses), $\gamma_{u}$, and the year fixed effects, $\delta_{t}$. The general unit of observation $u$ represents either a combination of product and destination or a firm-product combination. One step further, if we look at the second expression of Eq. \ref{eq: exogeneity}, we conclude that our errors are orthogonal after conditioning on the low-rank matrix, the fixed effects, and the treatment, $W_{ut}$. The equivalence between the first and the second expression of Eq. \ref{eq: exogeneity} means that the treatment is assumed orthogonal, i.e., there is no interference (or, similarly, no policies' spillovers) between treated and untreated observations. From this perspective, we can say that the model is robust to time-varying confounders.\\

Briefly, in our exercise, we can say that the time-varying information we obtain from the trade matrix, even after being reduced to a lower rank, and to the observation- and time-specific fixed effects, is sufficient to capture all the signals we need. If this is true, we can say the model is robust to time-varying confounders\footnote{Please note that matrix completion does not study autocorrelation, although it applies to panel data. Trend predictability rather depends on the length of the panels. Longer panels will capture trends better than shorter panels for the simple reason that we can work on a longer sequence of predictors for each observation in the present.}. 

Compared with a traditional difference-in-differences method, we do not need to impose a parallel-trends assumption, because observed units can follow different trends, and the model captures them. Specifically, we do not extract a control group from the untreated observations to derive a counterfactual. Our multidimensional counterfactuals are obtained as predictions, potentially with a weight assigned to each cell of the entire trade matrix. To avoid overfitting, the penalty will help reduce the rank of the matrix to a lower value while preserving sufficient information. 

Similarly to a traditional difference-in-differences method, we assume the Stable Unit Treatment Value Assumption (SUTVA), which implies that the treatment assigned to one unit does not affect the outcome of another unit (no interference) and that treatment effects are consistent (no hidden variation). The SUTVA assumption is often difficult to apply in the complex relationships underlying trade data and trade policies. Briefly, a violation of 'no interference' implies that the elimination of a tariff on a product would directly affect another product. On the other hand, a violation of 'no hidden variation' implies that the magnitude of the tariffs before and after the treaty signature is unrelated to the outcome variation. In the following paragraphs, we will introduce sensitivity and robustness checks to demonstrate that violations of SUTVA, albeit possible, would not fundamentally affect the validity of our results.

\section{Results}
\label{sec: analysis}
In this section, we discuss the findings of our application. We perform two exercises: one at the product-destination level, where we also introduce the notions of intensive and extensive margins; another at the firm-product level, which implies we select multiproduct firms with their product portfolios. In each case, we start by describing the specific design of the matrix structure that we draw before running the estimator. Then, we report the prediction accuracies always needed to validate the model. Finally, we comment on the results using a few post-estimation statistics.

\subsection{Product-level analysis}
\label{sec: product_level}
The unit of observation is the product $p$ at the 6-digit level of the HS classification exported at time $\tau$ to a destination $d$. A product is \textit{treated} if its tariff or quotas have changed after CETA since September 2017. Therefore, in this section, we evaluate idiosyncratic Treatment Effects on the Treated (TET) in percentage points, which we can write as $TET_{pdt}^*$, since the general $u$-th unit of observation introduced in Section \ref{sec: matrix_completion} is now represented by a product $p$, at destination $d$, and time $t$.

For our purpose, besides Canada, we aggregate and rank major destinations of French exports to avoid matrix sparsity.\footnote{As in \citet{fontagne2018exporters}, we also observe a high sparsity because the selection of products at each destination is stringent. In the original data, the vector of products exported to each destination contains, on average, at least 80\% of zeros. A highly sparse matrix with excessive zeros complicates calculations and saturates computer memory.} We compute two separate destination rankings, and then we consolidate them. At first, we rank importing countries by the average total trade value they received from France during 2012-2016. In a second exercise, we rank destinations after counting the number of products received from France in the same period. Finally, we include in our selection those countries that are in the top ten in either ranking. The remaining destinations are mainly aggregated by continent (e.g., the rest of Europe, the rest of Asia, etc.). In Appendix Table \ref{tab:dest_Aggr}, we record the relative trade importance of each destination in our final ranking. We end up with 16 destinations, including Canada and the rest of the world.

As for products, we ensure we can properly separate the intensive and the extensive margin. In the first case, we only consider the subset of products that were exported to Canada in either of the two years before the treatment and were still exported after the CETA.\footnote{For a visual representation of the trade patterns included in the intensive margin, see Appendix Table \ref{app_tab:product_selection}.}

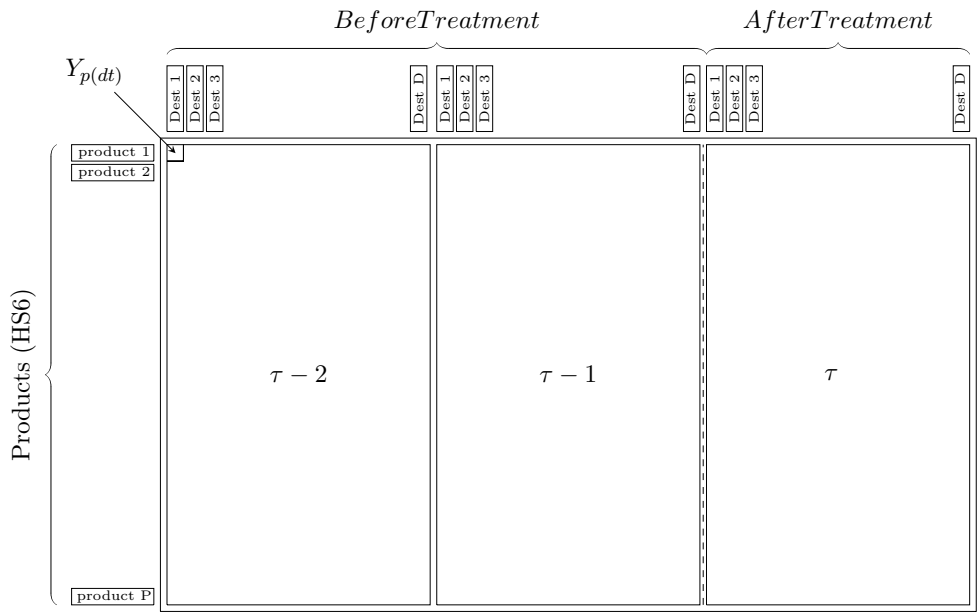
\begin{figure}
\centering
\caption{Matrix Structure for the product-level analysis}
    \label{fig:matrix_structure_intensive}
\resizebox{0.8\textwidth}{!}{
\centering
\begin{tikzpicture}
\draw (-0.1,-0.1) -- (12.3,-0.1) -- (12.3,7.1) -- (-0.1,7.1) -- (-0.1,-0.1);

\draw [dashed] (8.15,0) -- (8.15,7);


 \path 
  (2,3.5) node {\small  $\tau-2$}
  (6.1,3.5) node {\small $\tau-1$}
  (10.1,3.5) node {\small  $\tau$}
  ;
\draw (0,0) -- (4,0) -- (4,7) -- (0,7) -- (0,0);
\draw (4.1,0) -- (8.1,0) -- (8.1,7) -- (4.1,7) --(4.1,0);
\draw (8.2,0) -- (12.2,0) -- (12.2,7) -- (8.2,7) --(8.2,0);

 \path (0.125,7.7) node {\rotatebox{90}{\tiny Dest 1}}
  (0.425,7.7) node {\rotatebox{90}{\tiny Dest 2}}
  (0.725,7.7) node {\rotatebox{90}{\tiny Dest 3}};
        
\draw (0.0,7.2) -- (0.25,7.2) -- (0.25,8.2) -- (0.0,8.2) -- (0.00,7.2);
\draw (0.3,7.2) -- (0.55,7.2) -- (0.55,8.2) -- (0.3,8.2) -- (0.3,7.2);
\draw (0.6,7.2) -- (0.85,7.2) -- (0.85,8.2) -- (0.6,8.2) -- (0.6,7.2);

 \path (3.825,7.7) node {\rotatebox{90}{\tiny Dest D}};
  
\draw (3.7,7.2) -- (3.95,7.2) -- (3.95,8.2) -- (3.7,8.2) -- (3.7,7.2);

  \path (4.225,7.7) node {\rotatebox{90}{\tiny Dest 1}}
  (4.525,7.7) node {\rotatebox{90}{\tiny Dest 2}}
  (4.825,7.7) node {\rotatebox{90}{\tiny Dest 3}};
        
\draw (4.1,7.2) -- (4.35,7.2) -- (4.35,8.2) -- (4.1,8.2) -- (4.1,7.2);
\draw (4.4,7.2) -- (4.65,7.2) -- (4.65,8.2) -- (4.4,8.2) -- (4.4,7.2);
\draw (4.7,7.2) -- (4.95,7.2) -- (4.95,8.2) -- (4.7,8.2) -- (4.7,7.2);

  \path (7.955,7.7) node {\rotatebox{90}{\tiny Dest D}} ;
  
\draw (8.1,7.2) -- (7.85,7.2) -- (7.85,8.2) -- (8.1,8.2) -- (8.1,7.2);

  \path (8.325,7.7) node {\rotatebox{90}{\tiny Dest 1}}
  (8.625,7.7) node {\rotatebox{90}{\tiny Dest 2}}
  (8.925,7.7) node {\rotatebox{90}{\tiny Dest 3}};
        
\draw (8.2,7.2) -- (8.45,7.2) -- (8.45,8.2) -- (8.2,8.2) -- (8.2,7.2);
\draw (8.5,7.2) -- (8.75,7.2) -- (8.75,8.2) -- (8.5,8.2) -- (8.5,7.2);
\draw (8.8,7.2) -- (9.05,7.2) -- (9.05,8.2) -- (8.8,8.2) -- (8.8,7.2);

  \path (12.055,7.7) node {\rotatebox{90}{\tiny Dest D}};
  
\draw (12.2,7.2) -- (11.95,7.2) -- (11.95,8.2) -- (12.2,8.2) -- (12.2,7.2);

\path (-0.8,0.125) node {\tiny product P}
(-0.8,6.575) node {\tiny product 2}
(-0.8,6.875) node {\tiny product 1};
\draw (-0.2,0) -- (-1.45,0) -- (-1.45,0.25) -- (-0.2,0.25) -- (-0.2,0);
\draw (-0.2,6.75) -- (-1.45,6.75) -- (-1.45,7) -- (-0.2,7) -- (-0.2,6.75);
\draw (-0.2,6.7) -- (-1.45,6.7) -- (-1.45,6.45) -- (-0.2,6.45) -- (-0.2,6.7);

\draw (0,6.75) -- (0.25,6.75) -- (0.25,7) -- (0.25,6.75) -- (0,6.75);
 \draw [stealth-](0.13,6.87) -- (-0.8,7.8);
\path (-1.1,8.1) node {\small $Y_{p(dt)}$};
\draw [
    decorate, 
    decoration = {calligraphic brace,
        raise=5pt,
        amplitude=5pt,
        aspect=0.5}] (-1.5,0) --  (-1.5,7)
node[pos=0.5,left=10pt,black]{\rotatebox{90}{\small Products (HS6)}};

\draw [
    decorate, 
    decoration = {calligraphic brace,
        raise=5pt,
        amplitude=5pt,
        aspect=0.5}] (0,8.2) --  (8.2,8.2)
node[pos=0.5,above=10pt,black]{\small $Before Treatment$};

\draw [
    decorate, 
    decoration = {calligraphic brace,
        raise=5pt,
        amplitude=5pt,
        aspect=0.5}] (8.2,8.2) --  (12.2,8.2)
node[pos=0.5,above=10pt,black]{\small $After Treatment$};
\end{tikzpicture}
}
\end{figure}

In Figure \ref{fig:matrix_structure_intensive}, we visualize our matrix structure. For the intensive margin, the $P$ rows of the matrix correspond to the HS 6-digit products exported by France. The $TD$ columns of the matrix, instead, correspond to the set of $D$ possible export destinations in $T$ different times. Then, each matrix element $Y_{pdt}$ is the total export value for product $p$ at destination $d$ and time $t$. 

For the extensive margin, our focus is on the effect on the export probability of treated products. In this case, we will consider all possible products $\mathcal{P}$ exported by France anywhere, and each matrix element is a binary variable, $Y_{pdt}=\{0, 1\}$, which takes the value $1$ if product $p \in \mathcal{P}$ is exported at destination $d$ in time $t$, and $0$ otherwise. 

\begin{table}[htb]
   \centering
    \caption{Prediction accuracy at the product level}
    \label{tab: pred_quality_prod}
   \resizebox{.7\textwidth}{!}{
   \begin{tabular}{@{\extracolsep{5pt}} cccccc} 
\\[-1.8ex]\hline 
\hline \\[-1.8ex] 
  model & min RMSE & $\overline{Y}$ & SI & NRMSE & AUC\\ 
\hline \\[-1.8ex] 
Intensive Margin& 3.54310&7,624,650 &	0.00005& 0.00011 & \\
Extensive Margin & 0.23275&	&&0.23275 & 0.98\\ 
\hline \\[-1.8ex] 
\end{tabular} }
\begin{tablenotes}
     \singlespacing
    \footnotesize
    \item Note: The table reports standard measures of prediction accuracy. $\overline{Y}$ is the average trade of a line $p$ in a year for any destination $d$, and it is used to compute the normalized version of the RMSE and the Scatter Index (SI). The value of $\overline Y$ indicates the average predicted counterfactual in monetary values. On the extensive margin, no normalization is required, as the predicted outcomes are already in the range ${0, 1}$.
\end{tablenotes}
\end{table}

We estimate the model by solving the minimization problem described in the generic eq. \ref{eq: generic}, and we obtain two matrices of predicted outcomes: one for the intensive margin and one for the extensive margin. Then, crucially, Table \ref{tab: pred_quality_prod} reports some measures of the prediction accuracy computed after excluding the treated observations. Briefly, a certain level of prediction accuracy on the matrix, excluding treated observations, guarantees that our empirical model returns valid counterfactuals for the treated observations. If the predicted values of untreated observations are close to their observed values, we expect minimal bias when evaluating the policy's impact. 

As in a standard machine learning framework, the algorithm is first trained on different in-sample subsets and then evaluated on out-of-sample segments. In our specific case, the evaluation is made with a minimum average Root Mean Squared Error (RMSE) obtained after five random folds.\footnote{\label{footnote: cross-validation} Following the original procedure by \citet{athey2021matrix}, five random folds are used as cross-validation to derive the optimal $\lambda^*_Y$ of eq. \ref{eq: generic}. For each $\lambda_{Y}$, we train our model in-sample on each $k$-th random training subset, $\mathcal{O}_k \subset \mathcal{O}$, and we compute $\hat{\mathbf{Y}}(\lambda_{Y(k)}, \mathcal{O}_k)$. We then calculate the RMSE for each out-of-sample $k^{th}$ testing set. We pick the $\lambda_Y$ that minimizes RMSE, thereby achieving better prediction accuracy. Thus, Table \ref{tab: pred_quality_prod} reports the minimum average RMSE corresponding to the optimal $\lambda^*_Y$.}. Notably, we observe high prediction quality in both the intensive and extensive margins, as indicated by the very small Root Mean Squared Error (NRMSE) and Scatter Index (SI). For the intensive margin, the average difference between the predicted and observed values is $3.54$, and for the extensive margin, it is $0.23$. For the intensive margin, we obtain a very good value of the Area Under the Curve (AUC)\footnote{ The curve plots the true positive rate (sensitivity) against the false positive rate (1 - specificity) across all possible classification thresholds. The AUC is the scalar value representing the area beneath the ROC. A higher AUC indicates a model that is better at identifying positive cases while minimizing false ones}.

\subsubsection{Products' intensive margin}

\begin{figure}[H] 
    \centering
    \caption{Distribution of the idiosyncratic Treatment Effects on the Treated (TET) - intensive margin} \label{fig: TET Canada distribution graph}
     \resizebox{0.6\textwidth}{!}{%
\includegraphics[width=\textwidth]{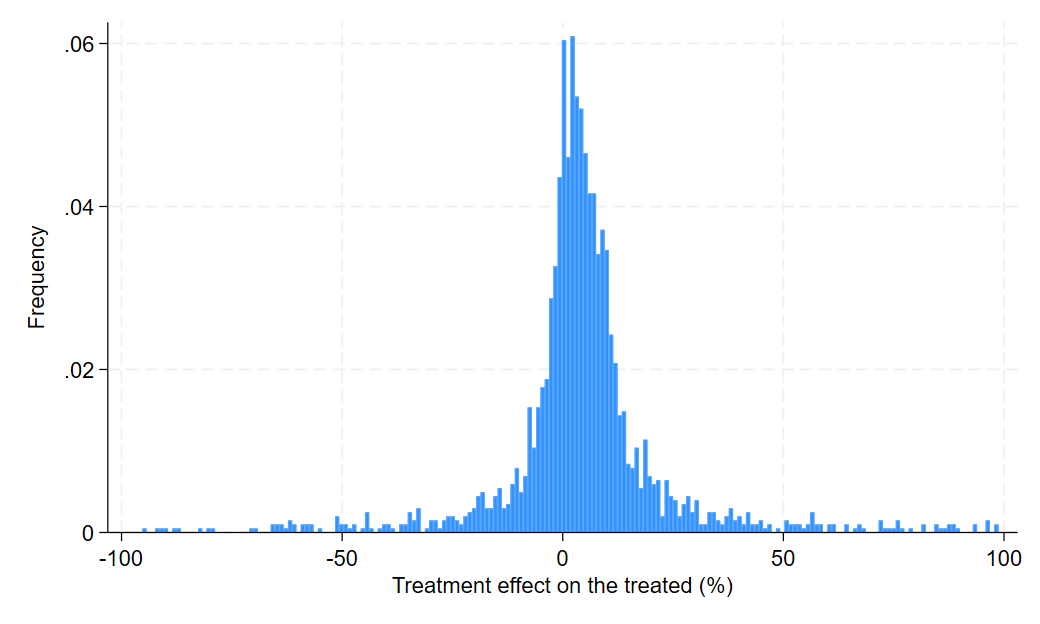}}
   \begin{tablenotes}
     \singlespacing
    \footnotesize
    \item Note: Relative treatment effects, $TET_{pdt}^*$, are computed following eq. \ref{eq: wATET} for each HS 6-digit product exported to Canada that has seen a change in the trade regime after CETA. 
\end{tablenotes}
\end{figure}

Let's start by looking at the heterogeneity of the treatment effects on the intensive margin for products exported to Canada in Figure \ref{fig: TET Canada distribution graph}, where we can observe both products that experienced a reduction in trade following the implementation of CETA, and products that consistently benefited from the new trade regime. The substantial heterogeneity in treatment effects needs special attention. We argue that exposing both positive and negative effects is an important advantage of implementing matrix completion, whereas the typical empirical test would summarize the policy's effectiveness with a single synthetic coefficient, therefore aggregating underlying heterogeneity. For example, when we implemented a simple diff-in-diff strategy (see Appendix B), we obtained a unique, statistically non-significant coefficient, leading us to conclude that the treaty had no impact. In reality, positive and negative effects can cancel each other out, and the unique coefficient can conceal relevant heterogeneity.

\begin{table}[!ht] 
\label{tab:results_prod}
    \centering
\caption{Weighted Treatment Effects on the Treated (wATET) products to Canada -  intensive margin} \label{tab: TET Canada distribution table}
     \resizebox{0.9\textwidth}{!}{%
    \begin{tabular}{lccccc}
    \hline
Model & Coeff. & St. error & St dev. & t-statistic & N. products \\
 & (1) & (2) & (3)\\ \hline
        ATET - Intensive margin & 5.531*** & 0.604 & 27.908& 9.1465 &2,221 \\ 
       weighted ATET - Intensive margin & 6.367 & - & 39.576 & - & 2,221 \\ \hline
    \end{tabular}}
    \begin{tablenotes}
    \footnotesize
    \item Note: The table reports first the Average Treatment Effects on the Treated products (ATET) exported in Canada after a t-test, and then a Weighted Average Treatment Effects on the Treated (wATET) products obtained from $TET_{pdt}^*$, considering each product's relevance in total exports of the year before the treaty signature. The standard deviation of wATET is computed as $\sqrt{\frac{\sum_{i=1}^N s_{pdt} \left( TET_{pdt}^*-wATET\right)^2}{(\mathcal L-1) \backslash \mathcal L \sum_{i=1}^N s_{pdt}}}$, where $\mathcal{L}$ is the number of counterfactuals in the trade matrix for Canada. \sym{***} stands for \(p<0.001\).
\end{tablenotes}
\end{table}

Visually, we can realize that positive treatment effects slightly prevail. In Table \ref{tab: TET Canada distribution table}, column (1), we report the weighted (wATET) and unweighted Average Treatment Effects on the Treated products (ATET), following eq. \ref{eq: wATET}, which can still be used to evaluate the overall impact of the new trade regime. We find an ATET of about 5.53\%, and a slightly higher wATET of 6.37\%. The statistical significance of ATET is determined by a t-test. For wATET we report a weighted standard deviation computed as $\sqrt{\frac{\sum_{i=1}^N w_{pdt} \left( TET_{pdt}^*-wATET_{pdt}\right)^2}{(\mathcal M-1) \backslash \mathcal M \sum_{i=1}^N w_{pdt}}}$, where we take into account the distribution of weights, $\mathcal{M}$ is the number of the treatment effects on the treated products that we computed, and $wATET_{pdt}$ is the weighted average we get from \ref{eq: wATET}.. Indeed, both in the case of the weighted and unweighted effects, the standard deviations are relatively high.

\begin{figure}
    \centering
    \caption{Distribution of the relative Treatment Effects (TE) on the intensive margin by main product classes} \label{fig: TET prod_classes graph}
     \resizebox{0.8\textwidth}{!}{%
\includegraphics[width=\textwidth]{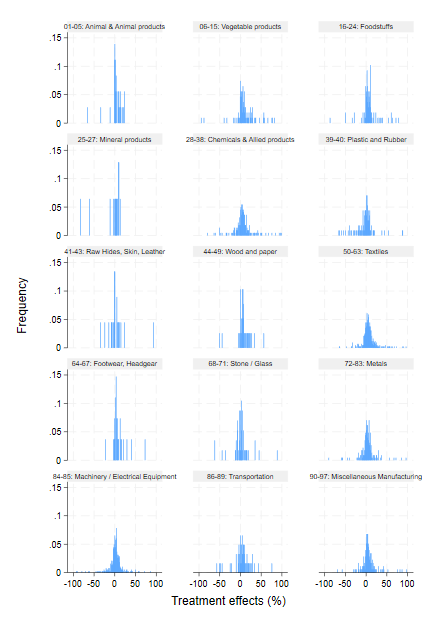}}
   \begin{tablenotes}
     \singlespacing
    \footnotesize
    \item Note: The figure reports histograms for the distribution by main product classes of relative treatment effects, $TE_{pdt^*}$, following eq. \ref{eq: wATET}, which have been computed for each HS 6-digit product exported to Canada that has seen a change in the trade regime after CETA, and then they are weighted for the relevance each product had in the year before the treaty signature.  
\end{tablenotes}

\end{figure}

\begin{table}[!ht]
    \centering
    \caption{Weighted Average Treatment Effects on the Treated (wATET) on exported products to Canada - intensive margin of main product classes} \label{tab: TET prod_classes}
         \resizebox{0.8\textwidth}{!}{%
    \begin{tabular}{clccc}
    \hline
      HS & Product class name & wATET & w. st. dev. & N. obs. \\ \hline 
        01-05 & Animal and Animal Products & 6.93 & 40.56 & 47 \\  
        06-15 & Vegetable Products & 9.18& 78.44 & 101 \\  
        16-24 & Foodstuffs & 8.31& 57.65 & 137 \\  
        25-27 & Mineral Products & 4.97 & 1.72 & 19 \\  
        28-38 & Chemicals and Allied & 6.01 & 25.76 & 249 \\  
        39-40 & Plastics and Rubber & 2.93 & 7.76 & 128 \\  
        41-43 & Raw Hides, Skins, Leather & 5.31 & 6.50 & 25 \\  
        44-49 & Wood and Paper & 6.56 & 15.13 & 25 \\  
        50-63 & Textiles & 8.41& 48.31 & 471 \\  
        64-67 & Footwear, Headgear & 6.17 & 27.37 & 31 \\  
        68-71 & Stone/Glass & 1.98 & 3.47 & 65 \\  
        72-83 & Metals & 8.68 & 71.22 & 249 \\  
        84-85 & Machinery/Electrical Equipment & 5.89 & 33.73 & 414 \\  
        86-89 & Transportation & 4.98 & 21.92 & 65 \\  
        90-97 & Miscellaneous Manufactured Articles & 5.62 & 29.29 & 195 \\  \hline    \end{tabular}}
    \begin{tablenotes}
    \footnotesize
    \item Note: The table reports the Weighted Average Treatment Effects on the Treated (wATET) exports by main product classes to Canada. Treatment effects in percentage points, $TET_{pdt}^*$, are weighted for each product's relevance in the year before the treaty signature to obtain the unique $wATET$. The weighted standard deviations are computed as $\sqrt{\frac{\sum_{i=1}^N s_{pdt} \left( TET_{pdt}^*-wATET\right)^2}{(\mathcal L-1) \backslash \mathcal L \sum_{i=1}^N s_{pdt}}}$, where $\mathcal{L}$ is the total number of the treatment effects on the treated units for the reference population of each row. .
\end{tablenotes}
\end{table}

The heterogeneity is still pronounced when we group single products by main HS classes, as in Table \ref{tab: TET prod_classes} and Figure \ref{fig: TET prod_classes graph}. Evidently, all product classes register a positive wATET. The impact is positive and higher in Vegetable Products, with a weighted average treatment effect (wATET) of 9.18\%, and lower in Stone and Glass Products, with a wATET of 1.98\%.

Nonetheless, when we evaluate the entire distribution of each product class, we always observe a fringe of products for which the signature of CETA has brought a negative impact. Even if such negative effects do not dominate the distributions, they remain relevant and warrant a few words' discussion. Why should we expect losses from trade agreements?

On the one hand, there is higher local competition. France was only the ninth-largest trading partner of Canada (See Section \ref{sec: introduction}), and a trade liberalization with the European Union implies that all member countries have access to the Canadian market on the same terms. In this context, it is only natural to observe that competition for French exporters to Canada has increased. If there had been a more favorable business environment for French producers in Canada, thanks to an albeit conflicted colonial past or a shared native language, it was reduced once other EU members' producers saw reduced barriers to entry into the market. 

On the other hand, there is underlying firms' selection dynamics. If we believe in \citet{bernard2007comparative}, we live in a world where firm heterogeneity à la \citet{melitz2003impact} interacts with Heckscher-Ohlin differences in endowments. In this case, we have that initial productivity differences magnify a country’s existing comparative advantage. A new trade agreement triggers the reallocation of resources and, through the selection of the most productive firms, raises average productivity, firm size, and job turnover faster in comparative advantage sectors. Total gains from trade are still positive, but the impact on sectors at a comparative disadvantage could be ambiguous. Against the previous background, let us introduce a few descriptive statistics that help qualify the positive and negative variations around the albeit positive average effect.

\begin{figure}
    \centering
    \caption{Treatment Effects on the Treated (TET \%) and comparative advantage - intensive margin} \label{fig: RCA}
     \resizebox{0.7\textwidth}{!}{%
\includegraphics[width=\textwidth]{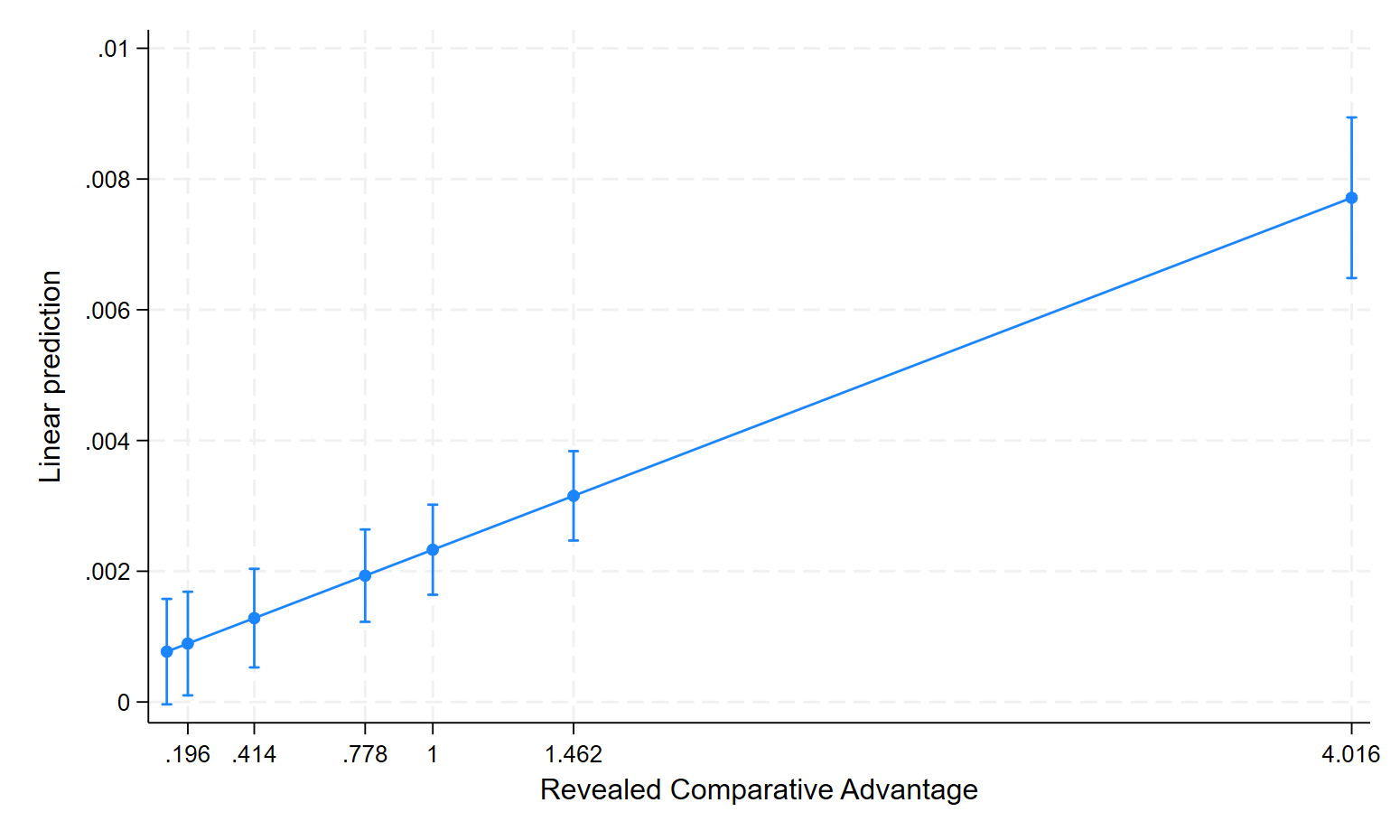}}
\begin{tablenotes}
     \singlespacing
    \footnotesize
    \item Note: The figure reports a plot of the predicted margins after a linear regression between the set of treatment effects on the treated in percentage points $TET_{pdt}^*$ when the destination is Canada and a standard measure of Revealed Comparative Advantage computed in the year before the CETA. The reference line, when RCA is equal to one, indicates that products below it were at a comparative disadvantage and products above it were at a comparative advantage. Bars indicate a 95\% confidence interval.
    \end{tablenotes}

\end{figure}

When we look at Figure \ref{fig: RCA}, we record a positive correlation between the treatment effects expressed as percentage points, $TET_{pdt}^*$, and a measure of revealed comparative advantage (RCA) computed in the year before treatment, considering the universe of French customs data.\footnote{The standard measure of revealed comparative advantage (RCA) that we compute is in the form: $RCA_{pt}=\frac{\frac{X_{CA,pt}}{X_{CA, t}}}{\frac{X_{W,pt}}{X_{W,t}}}$, where $X_{CA,pt}$ is the export flow of the single $p$ HS 6-digit product from France to Canada at time $t$, $X_{CA,t}$ is the total export to Canada at time $t$, $X_{W,pt}$ is the export of the same $p$ product from France to the world at time $t$, and finally $X_{W,t}$ is the total export from France at time $t$.} Eventually, in Figure \ref{fig: RCA}, we visualize the statistical association with a 95\% confidence interval. We observe that the correlation is positive and statistically significant above the threshold when RCA equals 1.

Briefly, Figure \ref{fig: RCA} shows that, as expected, the higher the previous comparative advantage of the product in Canada, the greater the positive impact of the CETA. When tariffs are reduced or quotas are extended, the response in percentage points is larger for products that were already selling well in the Canadian market. In a nutshell, a good portion of product-level heterogeneity in the effects of CETA is finally explained by initial comparative advantage positions.\footnote{The association is not statistically significant when the RCA is less than 1. In cases of products that were at a comparative disadvantage, when a product was not selling well in Canada, the impact is indeterminate.}

\subsubsection{Extensive margin}

\begin{figure}
    \centering
    \caption{Positive and negative extensive margin}
     \resizebox{0.6\textwidth}{!}{%
\includegraphics[width=\textwidth]{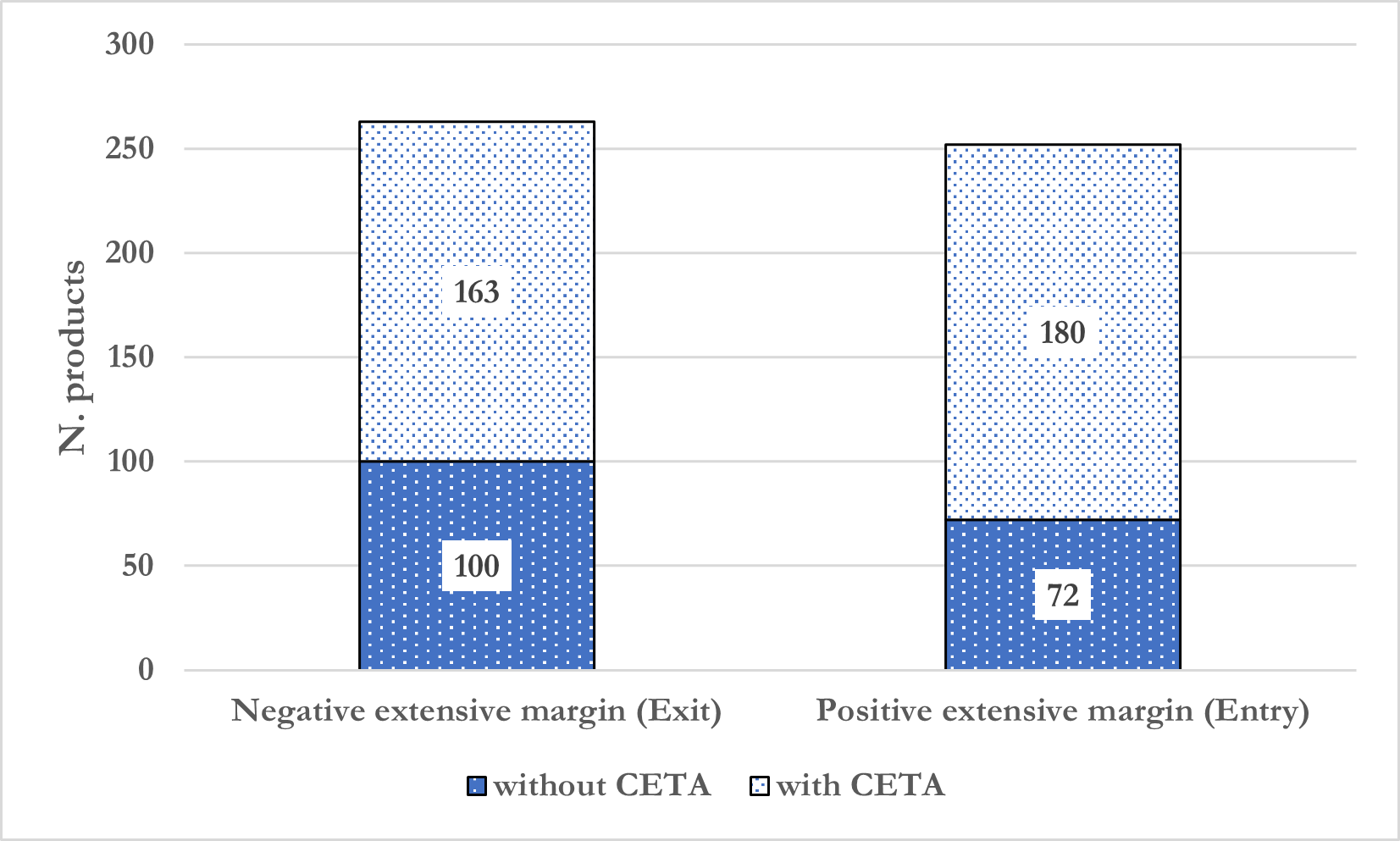}}
   \label{fig:prod_ext_mgn}
       \begin{tablenotes}
     \singlespacing
    \footnotesize
    \item Note: The figure reports the numbers of exiting (on the left) and entering products (on the right) that we observe after the signature of the CETA. The dark-coloured areas indicate products that would have entered or exited in any case, regardless of CETA, i.e., they are predicted as such in the matrix of potential outcomes. The light-colored area includes products that enter or exit Canada due to the CETA signature; i.e., they are assigned non-zero treatment effects after the potential-outcomes matrix. 
\end{tablenotes}
\end{figure}

Figure \ref{fig:prod_ext_mgn} provides a snapshot of the impact on the extensive margin, while corresponding numbers are reported in Table \ref{tab:prod_ext_mgn}. The impact is evaluated by considering the additional entry-exit dynamics due to CETA, on top of the regular entry-exit dynamics that we would have observed in any case, without the signature of the CETA. In Figure \ref{fig:prod_ext_mgn}, we start by separating the exiting products on the left and the entering products on the right. The dark-colored areas indicate, in both cases, the share of entry-exit that we do not attribute to the CETA because it is regularly predicted by our matrix as a counterfactual. The light-colored area instead represents cases of idiosyncratic treatment effects (TET) that are different from zero, as in eq. \ref{eq: tet}. If we compare with the number of incumbent products\footnote{We consider as incumbent the 2,221 products exported to Canada after the signature of the treaty and that were also exported at least once in the 5 years before the signature of the CETA. If we consider the demography predicted by the algorithm in the absence of the CETA, we would have about 3.2\% of regular entries and 4.4\% of regular exits. These numbers are close to what we find on aggregate in entry/exit in previous years, before CETA.} in 2017; the bar on the left indicates a positive impact on the extensive margin of about 180 products. That is, we registered an additional 180 products exported from France to Canada for the first time, thanks to CETA. On the other hand, we registered a negative impact on the extensive margin equal to 163 products. That is, we had an additional 163 products that were no longer exported due to CETA.

\begin{table}[!ht]
    \centering
    \caption{Positive and extensive margins - with and without CETA} \label{tab:prod_ext_mgn}
    \resizebox{0.7\textwidth}{!}{%
    \begin{tabular}{lccc}
    \hline
         & without CETA & with CETA & Total \\ \hline
        Negative extensive margin (Exit) & 100 & 163 & 263 \\ 
        Positive extensive margin (Entry) & 72 & 180 & 252 \\ \hline
    \end{tabular}}
    \begin{tablenotes}
     \singlespacing
    \footnotesize
    \item Note: The table reports the numbers of exiting (first row) and entering products (second row) that we observe after the signature of the CETA. In the first column, we report the number of products that have entered or exited due to the CETA, i.e., the non-zero treatment effects obtained from the matrix of potential outcomes. In the second column, we report the number of products that have entered or exited not due to CETA, i.e., those predicted to be so in the matrix of potential outcomes.
\end{tablenotes}
\end{table}

In Table \ref{tab: ext mgn produc classes}, we further separate negative and positive impacts on extensive margins by main HS product classes. We provide separate entries for entry-exit with and without the CETA (respectively w/ and w/o). Notably, the product class that has benefited the most from the treaty is Textiles (HS 50-63), with a net entry w/ CETA of 20, resulting from an entry w/ CETA of 37 and an exit w/ CETA of 17. It is followed by Vegetable Products (HS 06-15), with a net entry w/ CETA of 14, resulting from 25 new exported products vs. the exit w/ CETA of 11 products. Looking at the negative extensive margin, we find that the group with the highest number of exits due to CETA is Machinery/ Electrical (HS 84-85), with 26 products, followed by Chemicals (HS 28-38) with 25. The total net demography of exported products due to CETA is positive (17), while the regular net entry without the CETA has been negative (-28)

\begin{table}[H]
    \centering
    \caption{Extensive margin by main product classes}\label{tab: ext mgn produc classes}
     \resizebox{0.8\textwidth}{!}{%
    \centering
\begin{tabular}{clcc|cc|cc}
    \hline
        & & Entering&  &Exiting& & Net entry& \\
        HS & Product class & w/o & w/ & w/o & w/  & w/o & w/ \\ \hline
        01-05 & Animal \& Animal Products & 1 & 15 & 11 & 5 &-10&10\\ 
        06-15 & Vegetable Products & 5 & 25 & 16 & 11 &-11 &14\\ 
        16-24 & Foodstuffs & 1 & 5 & 4 & 8 &-3 & -3\\ 
        25-27 & Mineral Products & 1 & 5 & 1 & 6 &0 &-1\\ 
        28-38 & Chemicals \& Allied Products & 9 & 28 & 7 & 25 &2 &3 \\ 
        39-40 & Plastics / Rubbers & 2 & 5 & 1 & 5 &1 &0\\ 
        41-43 & Raw Hides, Skins, Leathers \& Furs & 0 & 0 & 4 & 1 &-4 &-1\\ 
        44-49 & Wood \& Wood Products & 4 & 7 & 11 & 11 &-7&-4\\ 
        50-63 & Textiles & 10 & 37 & 13 & 17 & -3 & 20\\ 
        64-67 & Footwear / Headgear & 1 & 0 & 0 & 2 & 1 & -2 \\ 
        68-71 & Stone / Glass & 4 & 4 & 3 & 10 &1 & -6\\ 
        72-83 & Metals & 8 & 18 & 8 & 23 & 0 & -5\\ 
        84-85 & Machinery / Electrical & 13 & 21 & 17 & 26 &-4 &-5\\ 
        86-89 & Transportation & 5 & 2 & 0 & 4 &5 &-2\\ 
        90-97 & Miscellaneous Manufacturing & 8 & 8 & 4 & 9 &4 &-1\\ \hline
        ~ & Total & 72 & 180 & 100 & 163 &-28 &17\\ 
        \hline
    \end{tabular}}
\begin{tablenotes}
     \singlespacing
    \footnotesize
    \item Note: The table reports the numbers of exiting, entering, and net entry of products by main HS product class, separated by whether they are due to CETA or they could have happened in any case, without the treaty signature. Entering and exiting products are identified as non-zero treatment effects in the potential outcomes matrix.
    \end{tablenotes}
\end{table}

\begin{figure}
    \centering
    \caption{Extensive margin and trade elasticity}
     \resizebox{0.7\textwidth}{!}{%
\includegraphics[width=\textwidth]{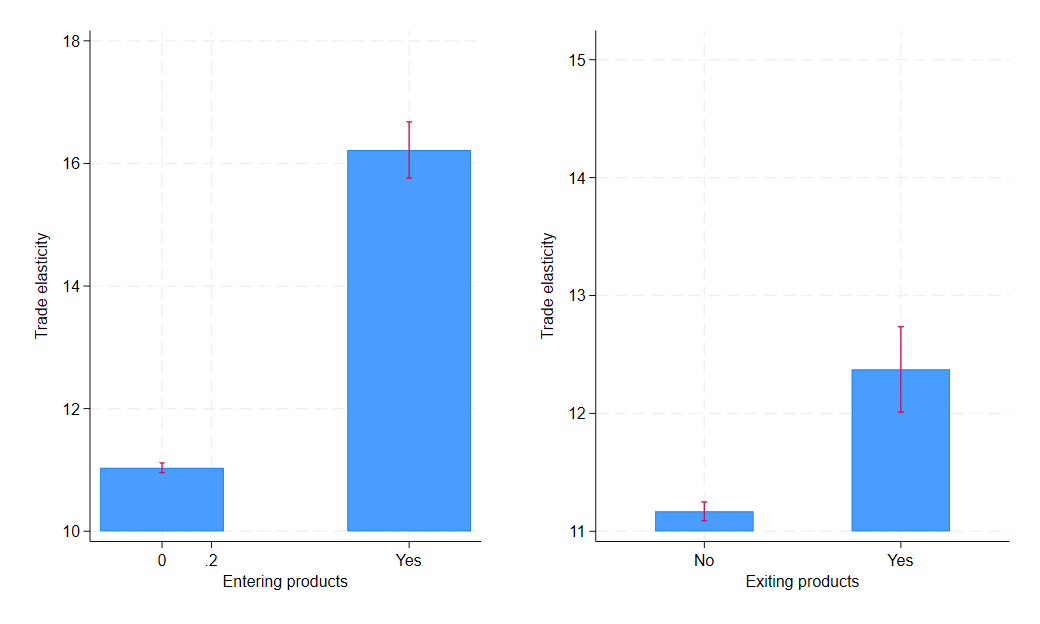}}
   \label{fig: Trade_elasticiy_EM}
\begin{tablenotes}
     \singlespacing
    \footnotesize
    \item Note: The figure shows the predicted margins from two linear probability models (LPMs), with the dependent variable being the trade elasticity of the single HS 6-digit product exposed to the CETA. In the left panel, the comparison is between the incumbent and the exiting products. In the right panel, the comparison is between the incumbent and the entering products. Trade elasticities are sourced from \cite{Fontagne_et_al_2022}. Bars indicate a 95\% confidence interval.
    \end{tablenotes}
\end{figure}

At this point, we can characterize the estimates of the extensive margin in Canada. Figure \ref{fig: Trade_elasticiy_EM} reports the results of two simple exercises. In both cases, we visualize the result of a linear regression model whose dependent variable is the trade elasticity of the single HS 6-digit product sourced from \cite{Fontagne_et_al_2022}. In the left panel, a binary variable (Yes/No) indicates whether the product entered the Canadian market under CETA or was already exported. On the right panel, a binary variable (Yes/No) indicates whether the product exited the Canadian market due to the CETA or continued to be available after the treaty. What we see is that entering and exiting products generally have higher trade elasticities than incumbent products. We believe it makes sense that products whose responses to changes in trade costs are relatively bigger are also the ones that react most to a tariff change, contributing to the extensive margin. In the case of a negative extensive margin, a fringe of exporters with relatively higher trade elasticities responds to changes in relative costs by reducing exports up to the point of exiting the Canadian market. Similarly, in the case of the positive extensive margin, a fringe of producers with relatively higher trade elasticities did not export to Canada and entered the market when they observed an albeit small tariff change.

\subsection{Firm-Level Analysis}
\label{sec: firm_level}

We choose to investigate the peculiar category of multi-product firms. The latter is an interesting category of firms that is certainly relevant, as we have seen in Figure \ref{fig:coverage firms}. which are responsible for about 85\% of export flows from France to Canada. From another perspective, multiproduct firms are also an interesting case to examine after trade liberalization events, as we can test whether they adjust their product portfolios as predicted by trade theory. 

From the original data, we select only those firms exporting more than one product to Canada within our time frame. Our baseline analyses will focus on portfolios with at least three products, while leaving the rest to sensitivity and robustness checks. Therefore, we generate a ranking for each firm by ordering products based on their export sales, from the most to the least sold in the year before the treaty signature. For this exercise, we will focus on the period 2015-2018.\footnote{Please note that an exercise on the intensive margins of product portfolios always needs two things: i) a fixed set of products per firm to properly design the trade matrix; ii) a proper separation from the different extensive margins (firms' entry-exit, products' entry-exit, entry-exit from the initial ranking). For this reason, we decided to focus on a shorter time period if confronted with the product-level intensive margin proposed in the previous paragraphs.} Notably, the first most traded products at the firm level account, on average, for 70\% of that firm's exports. 

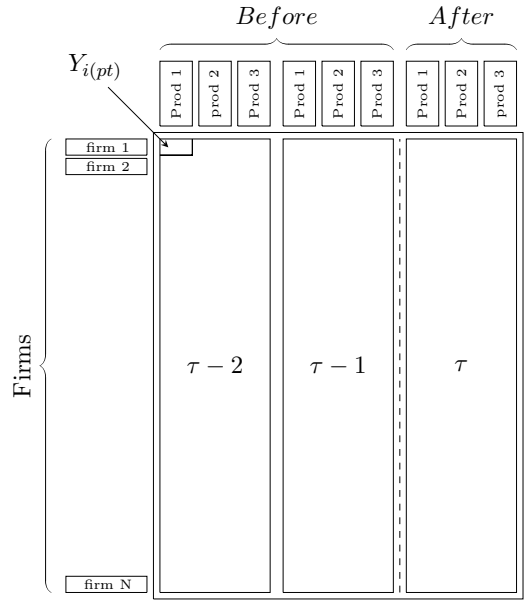
\begin{figure}
\centering
\caption{Matrix Structure for the firm-level analysis}
\label{fig:matrix_structure_intensive_firm}
\resizebox{!}{0.35\textheight}{
\centering
\begin{tikzpicture}
\draw (-0.1,-0.1) -- (5.6,-0.1) -- (5.6,7.1) -- (-0.1,7.1) -- (-0.1,-0.1);

\draw [dashed] (3.7,0) -- (3.7,7);


 \path 
  (0.85,3.5) node {\small  $\tau-2$}
  (2.75,3.5) node {\small $\tau-1$}
  (4.65,3.5) node {\small  $\tau$}
  ;
\draw (0,0) -- (1.7,0) -- (1.7,7) -- (0,7) -- (0,0);
\draw (1.9,0) -- (3.6,0) -- (3.6,7) -- (1.9,7) --(1.9,0);
\draw (3.8,0) -- (5.5,0) -- (5.5,7) -- (3.8,7) --(3.8,0);

 \path (0.25,7.7) node {\rotatebox{90}{\tiny Prod 1}}
  (0.85,7.7) node {\rotatebox{90}{\tiny prod 2}}
  (1.45,7.7) node {\rotatebox{90}{\tiny Prod 3}}
  ;
        
\draw (0.0,7.2) -- (0.5,7.2) -- (0.5,8.2) -- (0.0,8.2) -- (0.00,7.2);
\draw (0.6,7.2) -- (1.1,7.2) -- (1.1,8.2) -- (0.6,8.2) -- (0.6,7.2);
\draw (1.2,7.2) -- (1.7,7.2) -- (1.7,8.2) -- (1.2,8.2) -- (1.2,7.2);

  \path (2.15,7.7) node {\rotatebox{90}{\tiny Prod 1}}
  (2.75,7.7) node {\rotatebox{90}{\tiny Prod 2}}
  (3.35,7.7) node {\rotatebox{90}{\tiny Prod 3}}
  ;
        
\draw (1.9,7.2) -- (2.4,7.2) -- (2.4,8.2) -- (1.9,8.2) -- (1.9,7.2);
\draw (2.5,7.2) -- (3,7.2) -- (3,8.2) -- (2.5,8.2) -- (2.5,7.2);
\draw (3.1,7.2) -- (3.6,7.2) -- (3.6,8.2) -- (3.1,8.2) -- (3.1,7.2);

  \path (4.05,7.7) node {\rotatebox{90}{\tiny Prod 1}}
  (4.65,7.7) node {\rotatebox{90}{\tiny Prod 2}}
  (5.25,7.7) node {\rotatebox{90}{\tiny prod 3}}
  ;
        
\draw (3.8,7.2) -- (4.3,7.2) -- (4.3,8.2) -- (3.8,8.2) -- (3.8,7.2);
\draw (4.4,7.2) -- (4.9,7.2) -- (4.9,8.2) -- (4.4,8.2) -- (4.4,7.2);
\draw (5,7.2) -- (5.5,7.2) -- (5.5,8.2) -- (5,8.2) -- (5,7.2);

\path (-0.8,0.125) node {\tiny firm N}
(-0.8,6.575) node {\tiny firm 2}
(-0.8,6.875) node {\tiny firm 1}
;
\draw (-0.2,0) -- (-1.45,0) -- (-1.45,0.25) -- (-0.2,0.25) -- (-0.2,0);
\draw (-0.2,6.75) -- (-1.45,6.75) -- (-1.45,7) -- (-0.2,7) -- (-0.2,6.75);
\draw (-0.2,6.7) -- (-1.45,6.7) -- (-1.45,6.45) -- (-0.2,6.45) -- (-0.2,6.7);

\draw (0,6.75) -- (0.5,6.75) -- (0.5,7) -- (0.5,6.75) -- (0,6.75);
 \draw [stealth-](0.13,6.87) -- (-0.8,7.8);
\path (-1,8.1) node {\small 
$Y_{i(pt)}$};
\draw [
    decorate, 
    decoration = {calligraphic brace,
        raise=5pt,
        amplitude=5pt,
        aspect=0.5}] (-1.5,0) --  (-1.5,7)
node[pos=0.5,left=10pt,black]{\rotatebox{90}{\small Firms}};

\draw [
    decorate, 
    decoration = {calligraphic brace,
        raise=5pt,
        amplitude=5pt,
        aspect=0.5}] (0,8.2) --  (3.6,8.2)
node[pos=0.5,above=10pt,black]{\small $Before$};

\draw [
    decorate, 
    decoration = {calligraphic brace,
        raise=5pt,
        amplitude=5pt,
        aspect=0.5}] (3.8,8.2) --  (5.5,8.2)
node[pos=0.5,above=10pt,black]{\small $After$};
\end{tikzpicture}
}

\end{figure}

In Figure \ref{fig:matrix_structure_intensive_firm}, we report the design of a firm-level matrix to study the intensive margin by multiproduct firms. Please remember that, consistently with eq. \ref{eq: firm treatment}, we consider as treated any (multiproduct) firm with at least one product line whose tariff or quota has been affected by the signature of the CETA, while, in the following paragraphs, we will show the robustness to the inclusion of firms with all-treated products. 

In Figure \ref{fig:matrix_structure_intensive_firm}, rows correspond to the $N$ multi-product French exporters. Among them, $\Theta$ is the population of treated firms, and $(N-\Theta)$ is the set of untreated firms. Each column represents a different combination of time $t$ and product $p$. The product is identified at the HS 6-digit level, and we include only the three most traded lines for each firm before $\tau$, i.e., the year of treatment, among those exported in each of the three years in the panel. The matrix element $Y_{i,(pt)}$ measures the observed outcome of firm $i$ for the product $p$ at time $t$.

Similarly to what we did at the product level, we reconstruct the matrix of observed outcomes and predict the counterfactuals following the estimator in eq. \ref{eq: generic}. Table \ref{tab:pred_quality_firm_level} presents summary statistics of the prediction quality of our firm-level exercise. The percentage of expected error for the parameter of interest (i.e., the Scatter Index) is $29$\%. Prediction power indicates that the algorithm successfully replicates the dynamics of the original matrices of outcomes in the observed entries.\footnote{As in a classic machine-learning predictive framework, the algorithm is first trained on different in-sample subsets and then tested out of the sample. See also footnote \ref{footnote: cross-validation} for further details.} At this point, we can validly use predicted values of unobserved potential outcomes as counterfactuals for what would have happened if CETA were not signed.

\begin{table}[H]
\caption{Prediction quality - Firm-level analysis}
 \label{tab:pred_quality_firm_level}
        \centering
        \resizebox{.7\textwidth}{!}{
        \begin{tabular}{lccccc}
        \\[-1.8ex]\hline 
\hline \\[-1.8ex] 
           Model&n. obs.& $\overline{Y}$ &min Av(RMSE)&SI&NRMSE\\
           
    \midrule
           
            Intensive &3,177&203,345.61&59,069.2&29.04&42.93\\
                       \hline \\[-1.8ex] 
        \end{tabular}}
        \begin{tablenotes}
    \singlespacing
    \footnotesize
    \item Note: The table collects quality indicators for the predictions of observed values in the multiproduct firm-level exercise. The following columns indicate the average predicted value, the root mean squared error (RMSE), the scatter index, and the normalized RMSE.
\end{tablenotes}
    \end{table}

\subsubsection{Multiproduct firms and product scope}

\begin{figure}[H]
    \centering
    \caption{Distribution of treatment effects (\%) by product ranked in multiproduct firms}
     \resizebox{0.7\textwidth}{!}{%
\includegraphics[width=\textwidth]{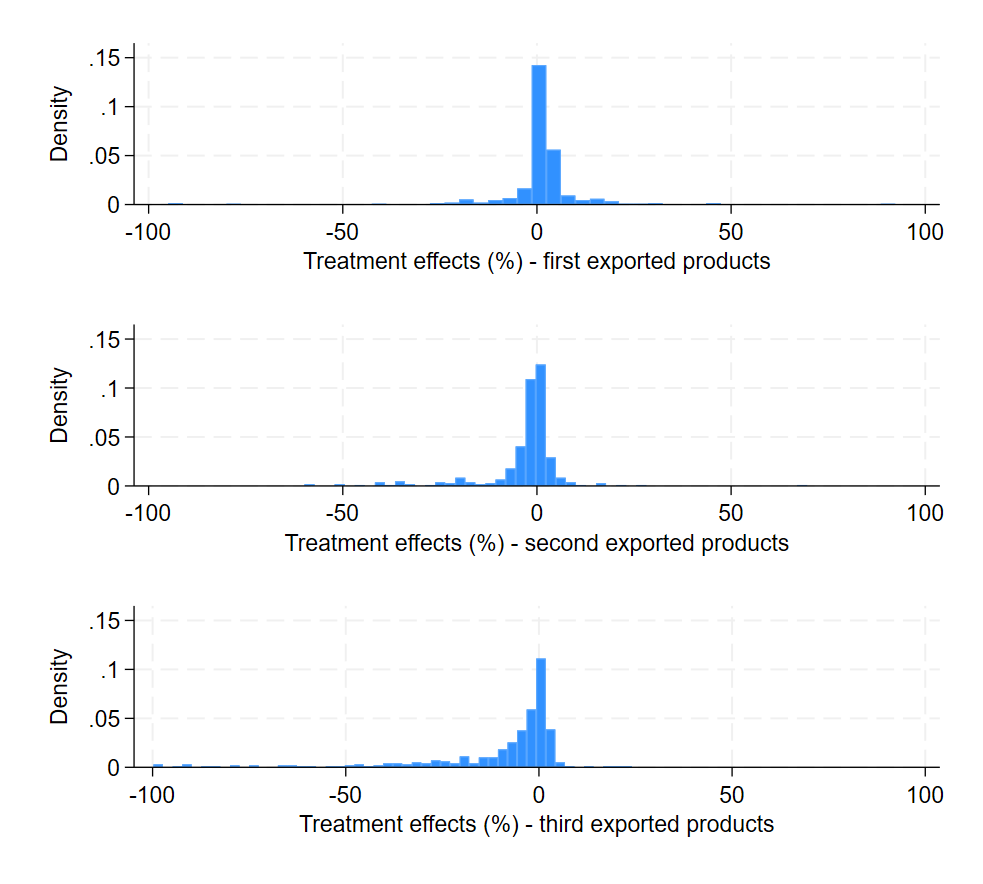}}
   \label{fig: Relative_TE_rank}
   \begin{tablenotes}
    \footnotesize
        \item Note: The Table shows the distribution of the treatment effects on the treated in percentage points, $TET_{ipt}^*$, for the first, second and third exported products in the multiproduct firms' portfolio.
    \end{tablenotes}
\end{figure}

Results on the impact of CETA on multiproduct firms are reported in Table \ref{tab: wATET multiproduct}, while Figure \ref{fig: Relative_TE_rank} reports a visualization of the distributions of treatment effects for the first, second, and third exported products, respectively. Please note that, in these paragraphs, we are considering multiproduct firms exposed to CETA and that exported at least 3 products in Canada, vs. a control group of untreated firms, as described in eq. \ref{eq: treated firms}. Therefore, our quantities of interest are the treatment effects on the treated, $TET_{ipt}^*$, expressed in percentage points with reference to products ordered, $p=\{1, 2, 3\}$, after considering their export sales in the firm's portfolio in the year before CETA. 

\begin{table}[!ht]
    \centering
    \caption{Weighted Average Treatment Effects on the Treated (wATET) products ranked by the multiproduct firms}
        \label{tab: wATET multiproduct}\resizebox{0.6\textwidth}{!}{%
    \begin{tabular}{lccc}
    \hline
        Type of firm/product & wATET & st. error & N. obs \\ \hline
         ~ & ~ & ~ & ~ \\
        \textit{\underline{All firms}} & ~ & ~ & ~ \\
        First exported product & 0.886* & 0.481 & 418 \\ 
        ~ & ~ & ~ & ~ \\ 
        Second exported product & 0.001 & 0.001 & 418 \\ 
        ~ & ~ & ~ & ~ \\ 
        Third exported product & 0.012*** & 0.001 & 418 \\ 
        ~ & ~ & ~ & ~ \\ 
        ~ & ~ & ~ & ~ \\ 
        \textit{\underline{Manufacturing firms}} &  &  & \\ 
        ~ & ~ & ~ & ~ \\ 
        First exported product & 0.729*** & 0.296 & 298 \\ 
        ~ & ~ & ~ & ~ \\ 
        Second exported product & -0.025*** & 0.001 & 298 \\ 
        ~ & ~ & ~ & ~ \\ 
        Third exported product & 0.001 & 0.001 & 298 \\ 
        ~ & ~ & ~ & ~ \\ 
        ~ & ~ & ~ & ~ \\ 
        \hline
    \end{tabular}}
    \begin{tablenotes}
    \footnotesize
        \item Note: The table reports the Weighted Average Treatment Effects on the Treated (wATET) exports for the first, second and third products in the multiproduct firms' portfolio. The $wATET$'s are computed considering products' trade shares in the year before the CETA. \sym{*}, \sym{**}, \sym{***} stand, respectively, for \(p<0.05\), \(p<0.01\),  \(p<0.001\).
    \end{tablenotes}
\end{table}

Looking at the first part of Table \ref{tab: wATET multiproduct}, we find that the weighted Average Treatment Effect on the Treated (wATET) first products for the sample of all exporters is 0.87\%, although weakly significant. At the same time, the wATET for the second product is not significantly different from zero, while the wATET on the third product indicates a tiny yet significant increase of $0.012\%$. Briefly, the CETA has, on average, a positive impact on at least two products out of three in the portfolio of multiproduct firms exposed to CETA. Yet, the impact is bigger for products already selling well in the Canadian market. Visually, our results are confirmed by the three graphs we included in Figure \ref{fig: Relative_TE_rank}, where, however, we can observe relevant heterogeneity distributed in the positive and negative quadrants.

Importantly, the second part of Table \ref{tab: wATET multiproduct} separates manufacturing firms from trade intermediaries, i.e., those firms that professionally act as intermediaries on behalf of other firms.\footnote{Originally, our data also included firms in primary markets, like agricultural products and other commodities, in the NACE rev. 2 sectors $01$-$09$. However, none of these firms are multiproduct according to the definition we introduced, and they are excluded from this part of the analysis.} Our separation is based on the NACE rev. 2 core activities of the firms, according to which we assume that wholesalers and retailers (NACE 45, 46 and 47) work as trade intermediaries in our data. It is interesting to see that, in the case of manufacturing firms, the wATET on the first exported products is about $0.73\%$, whereas the second exported products register an almost negligible $0.03\%$.

Not by chance, we believe previous results align with trade theory. We argue they are the product of a mechanism of portfolio adjustment, as in \cite{mayer2014market} and \cite{eckel2010multi}. According to previous literature, liberalization events lead to greater competition in export markets. In our case, more firms can access the Canadian market, and competitive pressure induces French exporters to concentrate their efforts on their best-performing products, thus focusing on their core competencies. Our findings are confirmed by a quick check on aggregate flows. According to our data, after trade liberalization between Canada and France with CETA, the first products by French exporters accounted for about 77\% of the total firms' exports. This number shows an increase compared to the 70\% share registered just before the treaty signature.

\section{Robustness and sensitivity checks}
\label{sec: robustness}

In this section, we discuss robustness or sensitivity checks on endogenous product selection, the simultaneity with the US-China trade wars, an alternative algorithmic choice, and a change in the definition of treatment on multiproduct firms.

Our first concern is that the parties have endogenously selected products during treaty negotiations, and we may mistakenly attribute a positive impact to selected products simply because they already showed higher trade potential. Clues of an endogenous selection into the treaty were offered in Table \ref{tab: endogenous selection}. Products in the CETA were already exported by a greater number of French firms, more frequently, with lower average transaction values and lower average value dispersion. Given our exogeneity assumption introduced in Section \ref{sec: exogeneity}, we cannot have anticipation effects, in the sense that the treatment cannot be correlated with the time-varying trade matrix.  
To address this concern, we conduct a placebo test by replicating the matrix completion analysis using the same definition of treated products as in the baseline, but for the period September 2012-August 2015. In Appendix Table \ref{appendix_tab:results_prod_placebo}, we report no significant ATET or wATET, which we argue provides supporting evidence for our empirical approach. Please see also Section \ref{sec: exogeneity} for a detailed discussion of the weaker exogeneity assumption on which matrix completion relies. Based on that assumption, we do not need to introduce specific robustness analysis, such as tests for parallel trends or the validity of a control group, because matrix completion does not rely on linear trends or on control groups.

\begin{figure}[H] 
    \centering
    \caption{Distribution of the idiosyncratic Treatment Effects on the Treated (TET) - products' intensive margin excluding products in US-China wars} \label{fig: TET Canada distribution graph no US-China wars}
     \resizebox{0.6\textwidth}{!}{%
\includegraphics[width=\textwidth]{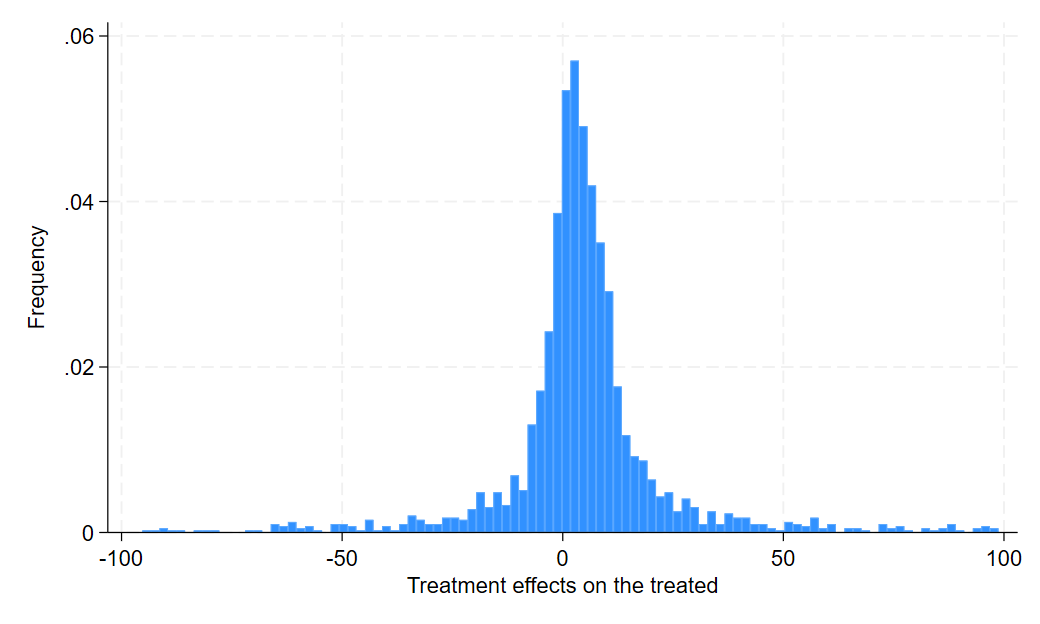}}
   \begin{tablenotes}
     \singlespacing
    \footnotesize
    \item Note: Relative treatment effects, $TET_{pdt}^*$, are computed following eq. \ref{eq: wATET} for each HS 6-digit product exported to Canada that has seen a change in the trade regime after CETA. Products included in the first stages of the US-China wars are excluded, as per \citet{Bown_2021}. The ATET is 4.43, the wATET is 5.88\%, and the number of products is 2,015.
\end{tablenotes}
\end{figure}

A second concern is that there may be a simultaneity bias arising from the co-occurrence of the CETA signature and the beginning of the US-China trade wars \footnote{Please note that the escalation of trade wars is subsequent to the period we test in this paper. According to \citet{Bown_2021}, by August 2018, about 9.8\% of Chinese exports were subject to US tariffs, and about 31\% of US exports were subject to Chinese tariffs. Only one year later, we observe that those numbers had increased to reach 47\% and 56\%, respectively.} For example, the positive results after CETA may in part be due to exports that were redirected from China and the US to Canada following the outbreak of the trade war\footnote{Similarly, we check that our results do not change after we exclude the softwood lumber (HS Code 440710), which had seen an increase in the tariff imposed by the US on Canada.}. In Figure \ref{fig: TET Canada distribution graph no US-China wars}, we report the distribution of idiosyncratic treatment effects after excluding a subset of HS 6-digit products extracted from a list provided by \citet{Bown_2021}. In particular, we exclude a set of 206 products that are affected by trade policy changes in the bilateral relationship between China and the US through July 2018. We find that the results are similar to the baseline: the ATET is 4.43\%, and the wATET is 5.88\%.

A third concern is that results depend on the specific choice of a matrix completion algorithm. As we discussed in Section \ref{sec: methods}, the main difference between the algorithm that we borrowed from \cite{athey2021matrix}, and other more standard proposals in computer science literature \citep{candes2010matrix, candes2012exact} is the inclusion of vectors of fixed effects before proceeding with the singular value decomposition. In our case, when we remove the vector of fixed effects, we find that the prediction performance worsens. We do not see a fundamental change in the results, so we prefer to keep our baseline.

Finally, we investigate what happens when we change the definition of treated firms. In our baseline, a multiproduct firm is treated if it exports at least two products to Canada and at least one of them is listed in the CETA. Briefly, by our definition, we have some treated firms with a portfolio that includes both products that have undergone a regime change and those that have not. If we change our definition to consider only firms that export all products listed in the CETA, we observe that the sample shrinks dramatically, becoming no longer representative. In fact, only 49 firms have three product all listed in CETA agreement; they are larger exporters, thus possibly introducing a major sample selection bias. We conclude that results with a different definition of treated firms cannot be trusted.

\begin{table}[!ht]
    \centering
    \caption{Weighted Average Treatment Effects on the Treated (wATET) products ranked by the multiproduct firms - only products with a trade regime change}
        \label{tab: robu multiproduct}\resizebox{0.6\textwidth}{!}{%
    \begin{tabular}{lccc}
    \hline
        Type of firm/product & wATET & st. error & N. obs \\ \hline
         ~ & ~ & ~ & ~ \\
        \textit{\underline{All firms}} & ~ & ~ & ~ \\
        First exported product & 25.356* & 12.181 & 49 \\ 
        ~ & ~ & ~ & ~ \\ 
        Second exported product & -12.015 & 15.008 & 49 \\ 
        ~ & ~ & ~ & ~ \\ 
        Third exported product & -0.012 & 7.006 & 49 \\ 
        ~ & ~ & ~ & ~ \\ 
        \hline
    \end{tabular}}
    \begin{tablenotes}
    \footnotesize
        \item Note: The table reports the Weighted Average Treatment Effects on the Treated (wATET) exports for the first, second and third products in the multiproduct firms' portfolio. We modify the definitions of treatment to include only firms that have all products with a regime change,.The $wATET$'s are computed considering products' trade shares in the year before the CETA. \sym{*} stand for \(p<0.05\).
    \end{tablenotes}
\end{table}

\section{Conclusions}
\label{sec: conclusion}

The present work proposes a novel approach to evaluating the impact of trade agreements using a causal machine learning framework. The aim is to provide a robust empirical strategy capable of handling the complexity and heterogeneity of trade effects at both the product and firm levels while mitigating concerns about endogenous selections into trade agreements. As a case study, we consider the entry into force of the EU-Canada Comprehensive Economic and Trade Agreement (CETA) and adapt the algorithm proposed by \citet{athey2021matrix} to the case of French customs data. The main advantage is that we can predict multidimensional counterfactuals at both the firm and product levels and, thus, obtain consistent estimates of causal effects. We also argue that matrix completion relies on a weaker exogeneity assumption than, for example, classical diff-in-diff exercises.

Our findings reveal a positive, average impact of the CETA on the product-level intensive margin in the year after the CETA. Yet, the product-level heterogeneity in the impact is relevant, and we show how matrix completion enables the evaluation of the full distribution of idiosyncratic treatment effects. Both positive and negative effects can be observed. Notably, they correlate with French comparative advantage (and disadvantage) distributions before the treaty. On the extensive margin, we observe product churn due to the treaty, which goes beyond the numbers of regular entry-exit dynamics. In line with expectations, entering and exiting products are also those that are more responsive to trade cost changes, i.e., whose trade elasticity is higher. 

Finally, at the firm level, we examine the case of multiproduct firms. Consistent with the mechanism of portfolio adjustment predicted by \citet{mayer2014market}, we observe that multiproduct exporters reallocate shares towards their first and most exported product, possibly due to an increasing local market competition after trade liberalization. Also in this case, we show how matrix completion allows evaluating an entire distribution of individual idiosyncratic treatment effects,

In conclusion, we believe that our matrix completion approach can be adapted to evaluate other trade policy actions, where multidimensional counterfactuals are relevant, as cells of a well-designed trade matrix. We argue that it returns a complete picture of the heterogeneity underlying trade regime changes.

\onehalfspacing
\setlength\bibsep{0.5pt}
\bibliographystyle{authordate3}
\bibliography{biblio.bib}

@article{yotov_2024,
  title={Staggered difference-in-differences in gravity settings: Revisiting the effects of trade agreements},
  author={Nagengast, Arne and Yotov, Yoto V},
journal={American Economic Journal: Applied Economics},
volume={},
number={},
pages={},
  year={2024}
}

@article{chaisemartin_2023,
  title={Two-way fixed effects and differences-in-differences with heterogeneous treatment effects: A survey},
  author={De Chaisemartin, Cl{\'e}ment and d’Haultfoeuille, Xavier},
  journal={The Econometrics Journal},
  volume={26},
  number={3},
  pages={C1--C30},
  year={2023},
  publisher={Oxford University Press}
}

@article{melitz_ottaviano_2008,
  title={Market size, trade, and productivity},
  author={Melitz, Marc J and Ottaviano, Gianmarco IP},
  journal={The review of economic studies},
  volume={75},
  number={1},
  pages={295--316},
  year={2008},
  publisher={Wiley-Blackwell}
}

@article{mrazova_2017,
  title={Not so demanding: Demand structure and firm behavior},
  author={Mr{\'a}zov{\'a}, Monika and Neary, J Peter},
  journal={American Economic Review},
  volume={107},
  number={12},
  pages={3835--3874},
  year={2017},
  publisher={American Economic Association 2014 Broadway, Suite 305, Nashville, TN 37203}
}

@incollection{goldberg_2016,
  title={The effects of trade policy},
  author={Goldberg, Pinelopi K and Pavcnik, Nina},
  booktitle={Handbook of commercial policy},
  volume={1},
  pages={161--206},
  year={2016},
  publisher={Elsevier}
}

@incollection{limao_2016,
  title={Preferential trade agreements},
  author={Lim{\~a}o, Nuno},
  booktitle={Handbook of commercial policy},
  volume={1},
  pages={279--367},
  year={2016},
  publisher={Elsevier}
}

@article{athey2021matrix,
  title={Matrix completion methods for causal panel data models},
  author={Athey, Susan and Bayati, Mohsen and Doudchenko, Nikolay and Imbens, Guido and Khosravi, Khashayar},
  journal={Journal of the American Statistical Association},
  pages={1--15},
  year={2021},
  publisher={Taylor \& Francis}
}

@article{candes2010matrix,
  title={Matrix completion with noise},
  author={Candes, Emmanuel J and Plan, Yaniv},
  journal={Proceedings of the IEEE},
  volume={98},
  number={6},
  pages={925--936},
  year={2010},
  publisher={IEEE}
}

@article{mazumder2010spectral,
  title={Spectral regularization algorithms for learning large incomplete matrices},
  author={Mazumder, Rahul and Hastie, Trevor and Tibshirani, Robert},
  journal={The Journal of Machine Learning Research},
  volume={11},
  pages={2287--2322},
  year={2010},
  publisher={JMLR. org}
}

@article{hubner2017eu,
  title={EU and trade policy-making: the contentious case of CETA},
  author={H{\"u}bner, Kurt and Deman, Anne-Sophie and Balik, Tugce},
  journal={Journal of European Integration},
  volume={39},
  number={7},
  pages={843--857},
  year={2017},
  publisher={Taylor \& Francis}
}

@article{fontagne2018exporters,
  title={Exporters’ product vectors across markets},
  author={Fontagn{\'e}, Lionel and Secchi, Angelo and Tomasi, Chiara},
  journal={European Economic Review},
  volume={110},
  pages={150--180},
  year={2018},
  publisher={Elsevier}
}

@article{melitz2003impact,
	author = {Melitz, Marc J},
	journal = {econometrica},
	number = {6},
	pages = {1695--1725},
	publisher = {Wiley Online Library},
	title = {The impact of trade on intra-industry reallocations and aggregate industry productivity},
	volume = {71},
	year = {2003}}

@article{baier2009estimating,
  title={Estimating the effects of free trade agreements on international trade flows using matching econometrics},
  author={Baier, Scott L and Bergstrand, Jeffrey H},
  journal={Journal of international Economics},
  volume={77},
  number={1},
  pages={63--76},
  year={2009},
  publisher={Elsevier}
}

@article{mayer2014market,
  title={Market size, competition, and the product mix of exporters},
  author={Mayer, Thierry and Melitz, Marc J and Ottaviano, Gianmarco IP},
  journal={American Economic Review},
  volume={104},
  number={2},
  pages={495--536},
  year={2014}
}

@article{eckel2010multi,
  title={Multi-product firms and flexible manufacturing in the global economy},
  author={Eckel, Carsten and Neary, J Peter},
  journal={The Review of Economic Studies},
  volume={77},
  number={1},
  pages={188--217},
  year={2010},
  publisher={Wiley-Blackwell}
}

@article{aitken1973effect,
  title={The effect of the EEC and EFTA on European trade: A temporal cross-section analysis},
  author={Aitken, Norman D},
  journal={The American Economic Review},
  volume={63},
  number={5},
  pages={881--892},
  year={1973},
  publisher={JSTOR}
}

@article{abrams1980international,
  title={International trade flows under flexible exchange rates},
  author={Abrams, Richard K},
  journal={Economic Review},
  volume={65},
  number={3},
  pages={3--10},
  year={1980},
  publisher={Federal Reserve Bank of Kansas City March}
}

@article{soloaga2001regionalism,
  title={Regionalism in the nineties: What effect on trade?},
  author={Soloaga, Isidro and Wintersb, L Alan},
  journal={The North American Journal of Economics and Finance},
  volume={12},
  number={1},
  pages={1--29},
  year={2001},
  publisher={Elsevier}
}

@article{bergstrand1985gravity,
  title={The gravity equation in international trade: some microeconomic foundations and empirical evidence},
  author={Bergstrand, Jeffrey H},
  journal={The review of economics and statistics},
  pages={474--481},
  year={1985},
  publisher={JSTOR}
}

@article{feenstra2001using,
  title={Using the gravity equation to differentiate among alternative theories of trade},
  author={Feenstra, Robert C and Markusen, James R and Rose, Andrew K},
  journal={Canadian Journal of Economics/Revue canadienne d'{\'e}conomique},
  volume={34},
  number={2},
  pages={430--447},
  year={2001},
  publisher={Wiley Online Library}
}

@techreport{baier2002endogeneity,
  title={On the endogeneity of international trade flows and free trade agreements},
  author={Baier, Scott L and Bergstrand, Jeffrey H},
  year={2002},
  institution={mimeo New York}
}

@article{baier2007free,
  title={Do free trade agreements actually increase members' international trade?},
  author={Baier, Scott L and Bergstrand, Jeffrey H},
  journal={Journal of international Economics},
  volume={71},
  number={1},
  pages={72--95},
  year={2007},
  publisher={Elsevier}
}

@article{magee2003endogenous,
  title={Endogenous preferential trade agreements: An empirical analysis},
  author={Magee, Christopher S},
  journal={Contributions in Economic Analysis \& Policy},
  volume={2},
  number={1},
  pages={1--17},
  year={2003},
  publisher={De Gruyter}
}

@article{head1998immigration,
  title={Immigration and trade creation: econometric evidence from Canada},
  author={Head, Keith and Ries, John},
  journal={Canadian journal of economics},
  pages={47--62},
  year={1998},
  publisher={JSTOR}
}

@article{YANG2014138,
title = {A panel data analysis of trade creation and trade diversion effects: The case of ASEAN–China Free Trade Area},
journal = {China Economic Review},
volume = {29},
pages = {138-151},
year = {2014},
issn = {1043-951X},
author = {Shanping Yang and Inmaculada Martinez-Zarzoso}}

@article{westerlund2011estimating,
  title={Estimating the gravity model without gravity using panel data},
  author={Westerlund, Joakim and Wilhelmsson, Fredrik},
  journal={Applied Economics},
  volume={43},
  number={6},
  pages={641--649},
  year={2011},
  publisher={Taylor \& Francis}
}

@article{bernard2007comparative,
  title={Comparative advantage and heterogeneous firms},
  author={Bernard, Andrew B and Redding, Stephen J and Schott, Peter K},
  journal={The Review of Economic Studies},
  volume={74},
  number={1},
  pages={31--66},
  year={2007},
  publisher={Wiley-Blackwell}
}

@article{bernard2010multiple,
  title={Multiple-product firms and product switching},
  author={Bernard, Andrew B and Redding, Stephen J and Schott, Peter K},
  journal={American economic review},
  volume={100},
  number={1},
  pages={70--97},
  year={2010},
  publisher={American Economic Association}
}

@article{bernard2011multiproduct,
  title={Multiproduct firms and trade liberalization},
  author={Bernard, Andrew B and Redding, Stephen J and Schott, Peter K},
  journal={The Quarterly journal of economics},
  volume={126},
  number={3},
  pages={1271--1318},
  year={2011},
  publisher={MIT Press}
}

@techreport{feenstra2007optimal,
  title={Optimal choice of product scope for multiproduct firms under monopolistic competition},
  author={Feenstra, Robert and Ma, Hong},
  year={2007},
  institution={National Bureau of Economic Research}
}

@incollection{baldwin2009impact,
  title={The impact of trade on plant scale, production-run length and diversification},
  author={Baldwin, John and Gu, Wulong},
  booktitle={Producer dynamics: New evidence from micro data},
  pages={557--592},
  year={2009},
  publisher={University of Chicago Press}
}

@article{abadie2010synthetic,
  title={Synthetic Control Methods for Comparative Case Studies:   Estimating the Effect of California's Tobacco Control Program},
  author={Abadie, Alberto and Diamond, Alexis and Hainmueller, Jens},
  journal={Journal of the American statistical Association},
  volume={105},
  number={490},
  pages={493--505},
  year={2010},
  publisher={Taylor \& Francis}
}

@techreport{arkhangelsky2019synthetic,
  title={Synthetic Difference in Differences},
  author={Arkhangelsky, Dmitry and Athey, Susan and Hirshberg, David A and Imbens, Guido W and Wager, Stefan},
  year={2019},
  institution={National Bureau of Economic Research}
}

@article{abadie2015comparative,
  title={Comparative Politics and the Synthetic Control Method},
  author={Abadie, Alberto and Diamond, Alexis and Hainmueller, Jens},
  journal={American Journal of Political Science},
  volume={59},
  number={2},
  pages={495--510},
  year={2015},
  publisher={Wiley Online Library}
}

@article{chernozhukov2021exact,
  title={An Exact and Robust Conformal Inference Method for Counterfactual and Synthetic Controls},
  author={Chernozhukov, Victor and W{\"u}thrich, Kaspar and Zhu, Yinchu},
  journal={Journal of the American Statistical Association},
  volume={116},
  number={536},
  pages={1849--1864},
  year={2021},
  publisher={Taylor \& Francis}
}

@article{candes2012exact,
  title={Exact matrix completion via convex optimization},
  author={Candes, Emmanuel and Recht, Benjamin},
  journal={Communications of the ACM},
  volume={55},
  number={6},
  pages={111--119},
  year={2012},
  publisher={ACM New York, NY, USA}
}

@article{qiu2013multiproduct,
  title={Multiproduct firms and scope adjustment in globalization},
  author={Qiu, Larry D and Zhou, Wen},
  journal={Journal of International Economics},
  volume={91},
  number={1},
  pages={142--153},
  year={2013},
  publisher={Elsevier}
}

@article{dhingra2013trading,
  title={Trading away wide brands for cheap brands},
  author={Dhingra, Swati},
  journal={American Economic Review},
  volume={103},
  number={6},
  pages={2554--84},
  year={2013}
}

@article{bas2013chinese,
  title={Chinese trade reforms, market access and foreign competition: The patterns of french exporters},
  author={Bas, Maria and Bombarda, Pamela},
  journal={the world bank economic review},
  volume={27},
  number={1},
  pages={80--108},
  year={2013},
  publisher={Oxford University Press}
}

@article{iacovone2010multi,
  title={Multi-product exporters: Product churning, uncertainty and export discoveries},
  author={Iacovone, Leonardo and Javorcik, Beata S},
  journal={The Economic Journal},
  volume={120},
  number={544},
  pages={481--499},
  year={2010},
  publisher={Oxford University Press Oxford, UK}
}

@article{BAIER200429,
title = {Economic determinants of free trade agreements},
journal = {Journal of International Economics},
volume = {64},
number = {1},
pages = {29-63},
year = {2004},
issn = {0022-1996},
author = {Scott L. Baier and Jeffrey H. Bergstrand}
}

@article{Rubin2005,
author = {Donald B Rubin},
title = {Causal Inference Using Potential Outcomes},
journal = {Journal of the American Statistical Association},
volume = {100},
number = {469},
pages = {322-331},
year = {2005},
publisher = {Taylor & Francis},

}

@article{Mullainathan_Spiess_2017,
Author = {Mullainathan, Sendhil and Spiess, Jann},
Title = {Machine Learning: An Applied Econometric Approach},
Journal = {Journal of Economic Perspectives},
Volume = {31},
Number = {2},
Year = {2017},
Month = {5},
Pages = {87-106},
}

@article{Athey_Imbens_2019,
author = {Athey, Susan and Imbens, Guido W.},
title = {Machine Learning Methods That Economists Should Know About},
journal = {Annual Review of Economics},
volume = {11},
number = {1},
pages = {685-725},
year = {2019},
}

@article{Angrist_Pischke_2010,
Author = {Angrist, Joshua D. and Pischke, Jörn-Steffen},
Title = {The Credibility Revolution in Empirical Economics: How Better Research Design Is Taking the Con out of Econometrics},
Journal = {Journal of Economic Perspectives},
Volume = {24},
Number = {2},
Year = {2010},
Month = {June},
Pages = {3-30}}

@article{Baier_et_al_2019,
title = {On the widely differing effects of free trade agreements: Lessons from twenty years of trade integration},
journal = {Journal of International Economics},
volume = {116},
pages = {206-226},
year = {2019},
issn = {0022-1996},
author = {Scott L. Baier and Yoto V. Yotov and Thomas Zylkin},
}

@misc{Larch_Yotov_2023,
  author       = {Larch, Mario and Yotov, Yoto V.}, 
  title        = {Estimating the Effects of Trade Agreements: Lessons from 60 Years of Methods and Data},
  howpublished = {CESifo Working Paper No. 10267},
  month        = 2,
  year         = 2023,
}

@article{Egger_Tarlea_2020,
author = {Egger, Peter H. and Tarlea, Filip},
title = {Comparing Apples to Apples: Estimating Consistent Partial Effects of Preferential Economic Integration Agreements},
journal = {Economica},
volume = {88},
number = {350},
pages = {456-473},
year = {2021}
}

@misc{Crowley_Han_2022,
  author       = {Crowley, Meredith and Han, Lu}, 
  title        = {The Pro-competitive Effects of Trade Agreements},
  howpublished = {CEPR Discussion Paper No. 17463},
  month        = 7,
  year         = 2022,
}

@article{Fontagne_et_al_2022,
title = {Tariff-based product-level trade elasticities},
journal = {Journal of International Economics},
volume = {137},
pages = {103593},
year = {2022},
author = {Lionel Fontagné and Houssein Guimbard and Gianluca Orefice},
}

@misc{Breinlich_et_al_2022,
  title        = {Machine Learning in International Trade Research - Evaluating the Impact of Trade Agreements},
author       = {Breinlich, H and Corradi V. and Rocha N. and Ruta M. and Santos Silva J. and Zylkin T.},
  howpublished = "{CEPR Discussion Paper No. 17325}",
  year         = 2022,
}

@article{Bartelme_et_al_2024,
title = {Specialization, market access and real income},
journal = {Journal of International Economics},
volume = {150},
pages = {103923},
year = {2024},
author = {Dominick Bartelme and Ting Lan and Andrei A. Levchenko},
}

@inproceedings{Athey_2015,
author = {Athey, Susan},
title = {Machine Learning and Causal Inference for Policy Evaluation},
year = {2015},
publisher = {Association for Computing Machinery},
address = {New York, NY, USA},
booktitle = {Proceedings of the 21th ACM SIGKDD International Conference on Knowledge Discovery and Data Mining},
pages = {5–6},
numpages = {2},
location = {Sydney, NSW, Australia},
series = {KDD '15}
}

@article{Knaus_2020,
    author = {Knaus, Michael C and Lechner, Michael and Strittmatter, Anthony},
    title = {Machine learning estimation of heterogeneous causal effects: Empirical Monte Carlo evidence},
    journal = {The Econometrics Journal},
    volume = {24},
    number = {1},
    pages = {134-161},
    year = {2020},
}

@article{Baiardi_Naghi_2024,
    author = {Baiardi, Anna and Naghi, Andrea A},
    title = {The value added of machine learning to causal inference: evidence from revisited studies},
    journal = {The Econometrics Journal},
    volume = {27},
    number = {2},
    pages = {213-234},
    year = {2024},
}

@misc{Poulos_et_al_2021,
    author = {Poulos, Jason and Albanese, Andrea and Mercatanti, Andrea and Fan, Li},
    title = {Retrospective Causal Inference via Matrix Completion, with an Evaluation of the Effect of European Integration on Cross-Border Employment},
    howpublished = {IZA Discussion Paper N. 14472},
    year = {2021},
}

@article{Xu_2017, 
title={Generalized Synthetic Control Method: Causal Inference with Interactive Fixed Effects Models}, 
volume={25}, 
number={1}, 
journal={Political Analysis}, 
author={Xu, Yiqing}, 
year={2017}, 
pages={57–76}}

@article{Abadie_2005,
    author = {Abadie, Alberto},
    title = {Semiparametric Difference-in-Differences Estimators},
    journal = {The Review of Economic Studies},
    volume = {72},
    number = {1},
    pages = {1-19},
    year = {2005},
}

@article{Bown_2021,
title = {The US–China trade war and Phase One agreement},
journal = {Journal of Policy Modeling},
volume = {43},
number = {4},
pages = {805-843},
year = {2021},
note = {The World Economy After Covid-19},
author = {Chad P. Bown},
}

\newpage

\appendix
\section*{Appendix A: Tables and graphs}\label{sec: appendix tables}

\setcounter{table}{0}
\renewcommand{\thetable}{A\arabic{table}}
\setcounter{figure}{0}
\renewcommand{\thefigure}{A\arabic{figure}}

\begin{table}[htpb!]
\caption{Distribution of tariff changes in the Canada-EU Comprehensive Trade Agreement (CETA) }
    \centering
    \begin{tabular}{ccc}
    \hline
        Tariff decrease (\%) & N. products & \% products \\ \hline
        0.3 - 5 & 1,871 & 51.04 \\ 
        6 - 10 & 1,290 & 35.19 \\ 
        11 - 20 & 479 & 13.06 \\ 
        $>$20 &     26 & 0.71 \\ \hline
        Total & 3,666 & 100.00 \\ \hline
    \end{tabular}
       \label{tab:tariff changes}
   \begin{tablenotes}
   \footnotesize
       \item Note: The table shows the distribution of tariff changes by HS 6-digit products as it has been negotiated in the CETA. The simple average tariff decrease has been 5.8\% with a 4.3 standard deviation.  \end{tablenotes}
\end{table}

\begin{figure}[htpb!]
    \centering
    \caption{Products per exporter in Canada in 2016}
     \resizebox{0.6\textwidth}{!}{%
\includegraphics[]{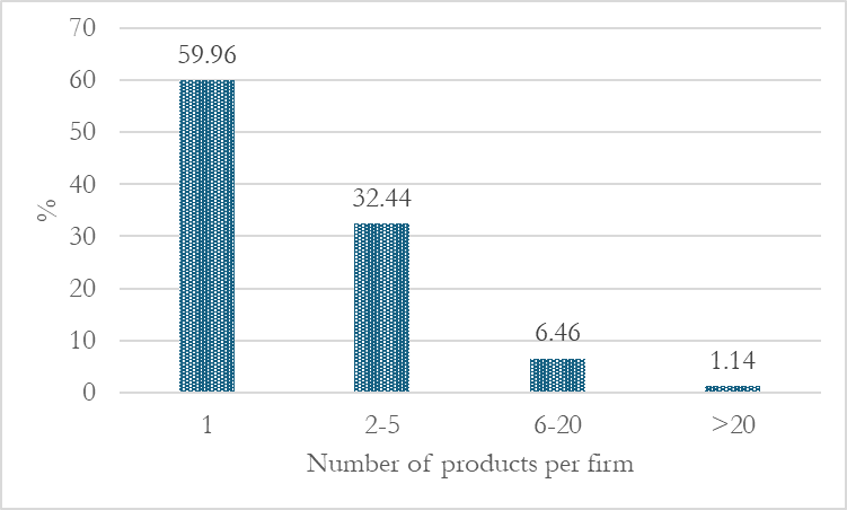}}
   \label{fig:products per exporter}
   \begin{tablenotes}
   \footnotesize
       \item Note: The figure shows the distribution of product portfolios by exporters to Canada before the entry into force of the CETA. On the left, the first bar indicates exporters with one product delivered to Canada. Then, the following bars refer to product portfolios sold to Canada by multiproduct firms. 
   \end{tablenotes}
\end{figure}

\begin{table}[H]
    \centering
        \caption{Which products in the intensive margin}
     \resizebox{0.8\textwidth}{!}{%
    \begin{tabular}{lccccp{6cm}}
    \hline\hline
    Case & Traded in & Traded in & Traded in & Intensive & Note:\\
    & 2015&2016&2017& margin&\\
    \hline
1)&Yes&Yes&Yes&Yes&Always traded\\
2)&Yes&Yes&No&No&Not traded after CETA\\
3)&Yes&No&Yes&Yes&Intermittently traded\\
4)&No&Yes&Yes&Yes&Intermittently traded\\
5)&Yes&No&No&No&Intermittently traded\\
6)&No&Yes&No&Yes&Intermittently traded\\
7)&No&No&Yes&No&Traded only after CETA\\
8)&No&No&No&No&Never traded\\
\hline
    \end{tabular}}
    \label{app_tab:product_selection}
    \begin{tablenotes}
    \footnotesize
        \item Note: The table separates cases of intensive margins from different trade patterns in the original data. For each of them, we report in column (4) whether the corresponding product is included in the analyses on intensive margins. 
    \end{tablenotes}
\end{table}

\begin{table}[htpb!]
    \centering
        \caption{Ranking export destination by trade volumes and number of products}
    \resizebox{0.8\textwidth}{!}{
\begin{tabular}{lccccc}
    \hline\hline
Destination &Export volume & \# Products & Rank by& Rank by& Combined \\
& (in mln €)&&Values&\#Products&Rank\\
& (1)&(2)&(3)&(4)&(5)\\
\hline
Germany&73.134&4,816&1&4&2.5\\
Italy&36.084&4,842&4&2&3\\
Spain&36.692&4,825&3&3&3\\
Belgium&30.752&4,857&6&1&3.5\\
USA&38.771&4,091&2&9&5.5\\
United Kingdom&35.721&4,594&5&7&6\\
Netherlands&16.350&4,775&8&5&6.5\\
Switzerland&15.922&4,691&9&6&7.5\\
China&19.489&3,836&7&10&8.5\\
Poland&8.356&4,193&10&8&9\\
Canada&4.217&3,812&26&18&22\\
\hline
Rest of Asia&74.958&4,890&&&\\
Rest of Europe&55.077&4,999&&&\\
Africa&29.825&4,912&&&\\
Rest of Americas&16.377&4,245&&&\\
Oceania&5.449&4,106&&&\\
\hline
\end{tabular}
}
    \label{tab:dest_Aggr}
    \begin{tablenotes}
    \singlespacing \footnotesize
        \item Note: Countries in this table are included in the trade matrix at the product level introduced in Section \ref{sec: product_level}. The decision is based on two criteria: in column (1), we report the average export sales by the French exporters from 2015-2016; in column (2), we report the average number of products exported to each destination in 2015-2016. Columns (3) and (4) report the ranking position of each country by average exports and average number of exported products, respectively. Column (5) reports an average of rankings in columns (3) and (4). The (rest of the) continents at the bottom of the table are also included in the analyses to close and balance the trade matrix.
    \end{tablenotes}
\end{table}

\begin{table}[htpb!]
\caption{Choice of destinations using different selection criteria}
    \label{appendix_tab: dest_selection_robustness}
    \centering
    \resizebox{0.8\textwidth}{!}{
    \begin{tabular}{lp{8cm}p{4cm}}
    \toprule
     	Selection Criterion&Individual Destinations&Aggregates\\
      \midrule
Baseline&Belgium, Canada, Switzerland, China, Germany, Spain, the United Kingdom, Italy, The Netherlands, Poland, the United States of America&Africa, Americas, Asia, Europe, Oceania\\
&&\\
Number of exporters&Belgium, Canada, Switzerland, China, Germany, Spain, the United Kingdom, Italy, Japan, Morocco, the United States of America&Africa, Americas, Asia, Europe, Oceania\\
&&\\
Import structure similarity&Austria, Australia, Canada, Germany, Spain, Finland, United Kingdom, New Zealand, Poland, Sweden, The United States of America&Africa, Americas, Asia, Europe, Oceania\\
&&\\
Import market size&China, Germany, the United Kingdom, Hong Kong, India, Italy, Japan, Korea, the Netherlands, the United States of America&Africa, Americas, Asia, Europe, Oceania\\
&&\\
\midrule
    \end{tabular}}
    \begin{tablenotes}\footnotesize
        \item \textit{Note:} The table reports, for each destination selection criterion, the list of partner countries included in the trade matrix. 
    \end{tablenotes}
\end{table}

 \begin{figure}[htpb!]
 \centering
 \caption{Time-destination correlation matrices on the extensive margins: original data matrix and the predicted low-rank matrix}
\label{fig:comp_models}
  \subfloat[Original data matrix $\mathbf{Y}$\label{fig: comp_models_original}]
 {\includegraphics[height=7cm,width=7cm]{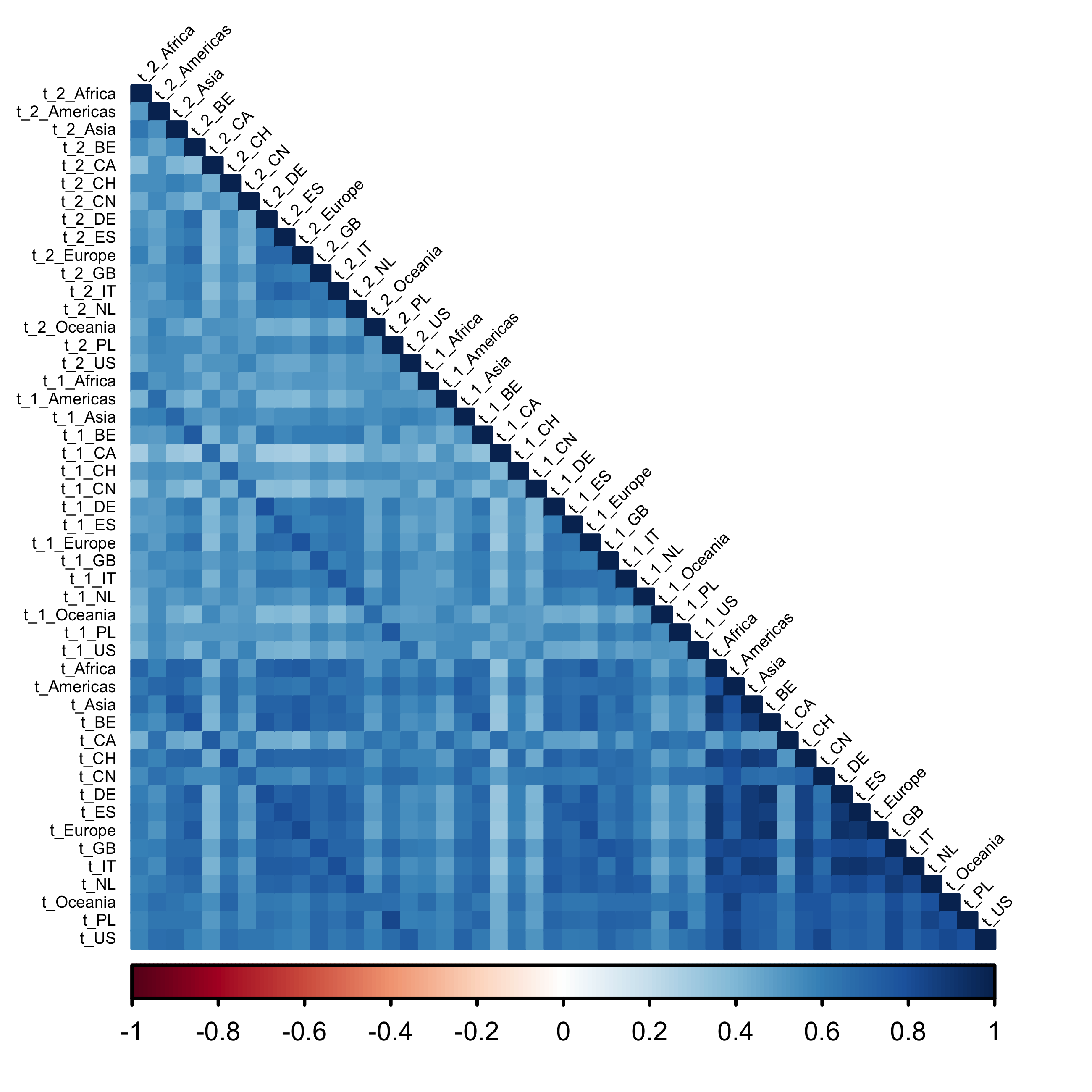}}\qquad
 \subfloat[Predicted matrix $\mathbf{\tilde Y}$\label{fig: comp_models_L}]
 {\includegraphics[height=7cm,width=7cm]{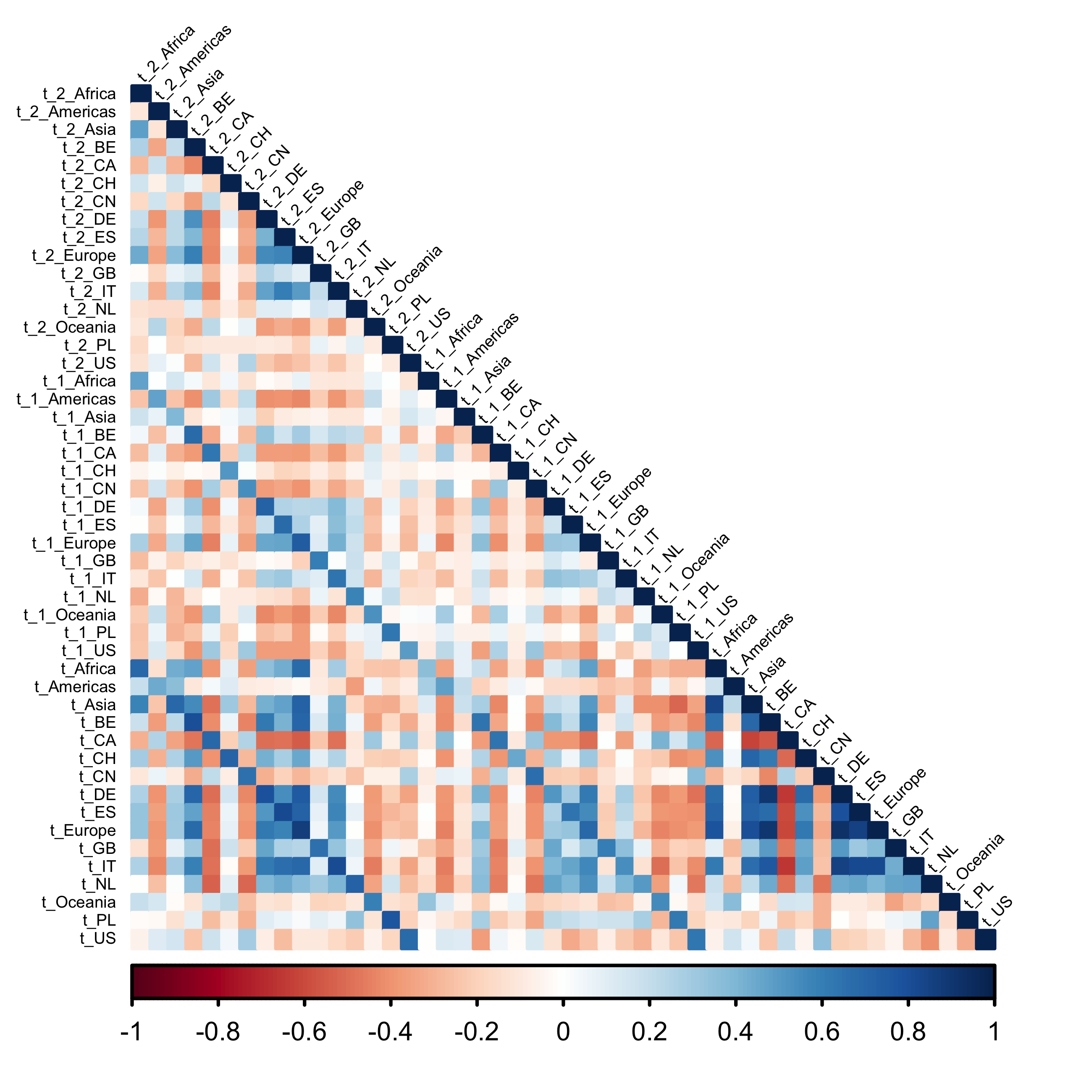}}
 \begin{tablenotes}
     \singlespacing
    \footnotesize
    \item Note: The figure on the left displays the correlation matrix of the columns of the original matrix $\mathbf{Y}$, which records the exports of products (at the HS6 level) to the country $d$ at time $t$. The figure on the right shows the correlation matrix of the columns of the corresponding predicted low-rank matrix $\mathbf{\tilde Y}$. This predicted matrix accounts for the residual correlation between the rows and columns of the original matrix after removing the row and column fixed effects ($\tilde \gamma_i$ and $\tilde \delta_j$ in equation \ref{eq: generic}). Figure (a) illustrates significant and consistent correlation patterns between destinations over time. Figure (b) demonstrates that these correlation patterns are effectively learned and captured by the predicted low-rank matrix $\mathbf{\tilde Y}$, enhancing the accuracy of the predicted outcomes. 
    \end{tablenotes}
\end{figure}

\begin{table}[htpb!] 

    \centering
\caption{A placebo test for the intensive margin to Canada} 
\label{appendix_tab:results_prod_placebo}
     \resizebox{0.7\textwidth}{!}{%
    \begin{tabular}{llccc}
    \hline
Product  &Class name & wATET & weighted st. dev. & N. products \\
 class&& (1) & (2) & (3) \\ \hline
        01-97&All products & -1.038& 11.664 & 2,219 \\ 
        &&&&\\
        01-05&Live animals \& Animal products&0.932&85.550&44\\
06-15&Vegetable products&5.380&0.696&122\\
16-24&Foodstuffs&0.415&4.262&120\\
25-27&Mineral products&-32.675&232.346&23\\
28-38&Chemicals \& Allied industries&-1.613&12.084&244\\
39-40&Plastics / Rubbers&-1.289&9.967&129\\
41-43&Raw Hides, Skins, Leather \& Furs&-1.021&5.609&31\\
44-49&Wood \& Wood products&0.578&9.423&31\\
50-63&Textiles&15.36&13.84&458\\
64-67&Footwear / Headgear&3.189&26.784&30\\
68-71&Stone / Glass&3.388&33.419&74\\
72-83&Metals&2.216&3.766&234\\
84-85&Machinery / Electrical&-1.655&4.607&418\\
86-89&Transportation&-9.612&6.021&66\\
90-97&Miscellaneous&1.253&3.382&195\\
        \hline
    \end{tabular}}
    \begin{tablenotes}
    \footnotesize
    \item Note: The table reports the Weighted Average Treatment Effects on the Treated (wATET) exports to Canada after a placebo test, considering the same definitions of treatment but in the period September 2012-August 2015. $TET_{pdt}^*$, are weighted for the relevance each product had in the year before the treaty signature to obtain the unique $wATET$. The weighted standard deviations are computed as $\sqrt{\frac{\sum_{i=1}^N s_{pdt} \left( TET_{pdt}^*-wATET\right)^2}{(\mathcal L-1) \backslash \mathcal L \sum_{i=1}^N s_{pdt}}}$, where $\mathcal{L}$ is the number of counterfactuals in the trade matrix for Canada. 
\end{tablenotes}
\end{table}

\begin{table}[htpb!]
\caption{Prediction accuracy at the product level intensive margin - Robustness checks}
    \centering
    \resizebox{0.7\textwidth}{!}{
    \begin{tabular}{lcccc}
    \toprule
     	Model&min RMSE&$\overline{Y}$	&SI	&NRMSE\\
      \midrule
&&&&\\
No fixed effects&	7.328702&	7,060,711&	0.000103796	&0.00027963\\
\midrule
    \end{tabular}}
    \label{appendix_tab: pred_quality_robustness}
    \begin{tablenotes}\footnotesize
        \item \textit{Note:} The table reports the statistics of the prediction accuracy that we obtain when we train the model while removing the fixed effects.
    \end{tablenotes}
\end{table}

\newpage

\section*{Appendix B: Difference-in-difference}\label{sec: DID}
\setcounter{table}{0}
\renewcommand{\thetable}{B\arabic{table}}
\setcounter{figure}{0}
\renewcommand{\thefigure}{B\arabic{figure}}

\onehalfspacing

We consider the simple difference-in-difference as a conventional empirical method for benchmarking against our preferred empirical strategy. According to our definitions, a treated product is a product listed in the CETA, while a treated firm is a firm that exports at least one product under CETA to Canada. Basic formulations are, for the intensive margins:

\begin{equation}
Y_{ut} = c_u + \gamma_t + \beta_D \cdot D_{ut} + \epsilon_{ut} \label{eq:diff-in-diff_int}
\end{equation}

the extensive margin for products:

\begin{equation}
    Pr(Q_{pt}=1|X_{pt}=1) = c_u + \gamma_t + \beta_D \cdot D_{ut} + \epsilon_{ut} \label{eq:diff-in-diff_ext_p}
\end{equation}

and the extensive margin for firms:

\begin{equation}
Q_{it}=\exp({c_{i}+\gamma_{t}+\beta_{D}D_{it}+\epsilon_{it}}) \label{eq:diff-in-diff_ext_i}
\end{equation}

where $Y_{ut}$ represents the total exports of the $u$-th unit of observation where $u = (p,i)$ is either a $p$-th product or an $i$-th-firm observed at time $t$ in Canada. Product fixed effects, $c_p$, and time fixed effects, $\gamma_t$, are included. The binary variable $D_{pt}$ is the treatment indicator, while the error term $\epsilon_{pt}$ captures stochastic variation. In eq. \ref{eq:diff-in-diff_ext_p}, we examine the impact of CETA on the product's extensive margin of trade with either a linear probability model (LPM) or a logit, whose dependent variable, $Q_{pt}$ is equal to one if the product was exported and zero otherwise. In eq. \ref{eq:diff-in-diff_ext_i}, instead, we study the impact of CETA on the firms' extensive margin with either a simple OLS or a Pseudo-Poisson, where $Q_{it}$ is the number of products exported in Canada by a firm $i$ at time $t$.

\begin{table}[htpb!]
    \centering
    \caption{Difference-in difference for products and firms}
    \label{tab:diff_in_diff}
\def\sym#1{\ifmmode^{#1}\else\(^{#1}\)\fi}
\resizebox{\textwidth}{!}{
\begin{tabular}{lccc|ccc}
\hline\hline
&\multicolumn{3}{c}{Product-level}&\multicolumn{3}{c}{Firm-level}\\
[0.5em]
&Intensive Margin&\multicolumn{2}{c}{Extensive Margin}&Intensive Margin&\multicolumn{2}{c}{Extensive Margin}\\
& (OLS) & (LPM) & (Logit)  & (OLS) &(OLS)&(Poisson)\\
&$Y_{pt}$&$P(Q_{pt}=1)$&$OR(Q_{pt}=1)$&$Y_{it}$&$Q_{it,CA}$&$Q_{it,CA}$\\
[.1em]
 & (1) & (2) & (3) & (4) & (5) &(6)\\ 
 [.5em]
\hline
ATT &91.63&0.017&1.244&-32.03&0.324\sym{***}&0.0842\sym{***}\\
&(115.1)&(0.010)&(0.161)&(48.52)&(0.075)&(0.028)\\
[1em]
Year fixed effects:&&&&&&\\
t-4&12.95&0.005&1.067&-5.452&-0.16&-0.008\\
&(46.53)&(0.005)&(0.085)&(16.65)&(0.030)&(0.010)\\
[1em]
t-3&79.97&0.016\sym{**}&1.228\sym{**}&21.00&0.127\sym{***}&0.040\sym{*}\\
&(58.09)&(0.005)&(0.085)&(16.65)&(0.049)&(0.016)\\
[1em]
t-2&15.83&0.014\sym{*}&1.201\sym{*}&-10.05&0.245 \sym{***}&0.077\sym{***}\\
&(62.16)&(0.006)&(0.087)&(25.52)&(0.061)&(0.020)\\
[1em]
t-1&89.88&0.033\sym{***}&1.526\sym{***}&18.91&0.420\sym{***}&0.133\sym{***}\\
&(62.61)&(0.006)&(0.113)&(22.83)&(0.070)&(0.021)\\
[1em]
t&82.20&0.015&1.219&76.37&0.232\sym{***}&0.085\sym{*}\\
&(128.9)&(0.010)&(0.150)&(50.76)&(0.051)&(0.023)\\
[1em]
constant&1,063.6\sym{***}&0.550\sym{***}&&291.1\sym{***}&2.407\sym{***}&\\
&(45.67)&(0.004)&&(16.54)&(0.043)&\\
[1em]
\hline
Product fixed effect&YES&YES&YES&NO&NO&NO\\
Firm fixed effect&NO&NO&NO&YES&YES&YES\\
N. obs.&15,763&31,236&10,980&53,338&53,338&45,729\\
\hline\hline
\end{tabular}}
\begin{tablenotes}
\singlespacing
\footnotesize
    \item Note: We report product-level results in columns 1-3. Column (1) reports results on the intensive margin expressed in thousands of euros. Columns (2) and (3) report results on the extensive margin either computed using a Linear Probability model (LPM) or a logit (Logit). We report firm-level results in columns 4-6. Column (4) reports the results on the intensive margin expressed in thousands of euros. Column (5) reports the results on the extensive margin computed using a Linear Model, while column (6) reports results after using a Pseudo-Poisson Model. Robust standard errors in parentheses. \sym{*}, \sym{**}, \sym{***} stand, respectively, for \(p<0.05\), \(p<0.01\),  \(p<0.001\).
\end{tablenotes}
\end{table}

\newpage
\section*{Appendix C: Prediction accuracy}

\setcounter{table}{0}
\renewcommand{\thetable}{C\arabic{table}}
\setcounter{figure}{0}
\renewcommand{\thefigure}{C\arabic{figure}}

\onehalfspacing

Different metrics are used to evaluate the prediction accuracy of machine learning algorithms. Briefly, prediction accuracy metrics compare the classes predicted by the algorithm with the actual ones. 

In the case of continuous outcomes, we can use the following measures:
\begin{itemize}
        \item \textbf{Root-Mean-Square Error (RMSE)}, which is computed as
        \begin{equation}
            RMSE=\sqrt{\sum_{i=1}^{NRD} (\hat{y}_{ird}-y_{ird})^2/NRD}
        \end{equation}
        \item  \textbf{Scatter Index (SI)}, computed as 
        \begin{equation}
            SI=RMSE/\overline{y}_{ird}*100
        \end{equation}
        It gives the percentage of expected error for the parameter of interest
        \item \textbf{Normalised Root-Mean-Square Error (NRMSE)}, computed as
        \begin{equation}
            NRMSE=RMSE/(Q3-y_{min})*100
        \end{equation}
        it relates the RMSE to the observed range of the variable, thus allowing comparisons with other models
    \end{itemize}

\end{document}